\documentclass[twocolumn]{aastex62} 

\usepackage{xspace}
\usepackage{color}
\usepackage{rotating}
\usepackage{amsmath}
\usepackage{subfigure}
\usepackage[version=4]{mhchem}
\usepackage[ampersand]{easylist}

\PassOptionsToPackage{hyphens}{url}
\usepackage{hyperref}
\usepackage{natbib}

\begin{document}

\bibliographystyle{apj}


\title{High-Resolution Spectral Discriminants of Ocean Loss for M Dwarf Terrestrial Exoplanets}

\shorttitle{High-Res Ocean Loss}
\shortauthors{Leung, Meadows, \& Lustig-Yaeger}

\correspondingauthor{Michaela Leung}
\email{mwjl@uw.edu}

\author[0000-0003-1906-5093]{Michaela Leung}
\affiliation{Department of Earth and Space Sciences, University of Washington, Seattle, Washington 98195, USA}
\affiliation{NASA Astrobiology Institute's Virtual Planetary Laboratory, Box 351580, University of Washington, Seattle, Washington 98195, USA}

\author[0000-0002-1386-1710]{Victoria S. Meadows}
\affiliation{Department of Astronomy and Astrobiology Program, University of Washington, Box 351580, Seattle, Washington 98195, USA}
\affiliation{NASA Astrobiology Institute's Virtual Planetary Laboratory, Box 351580, University of Washington, Seattle, Washington 98195, USA}
\affiliation{Astrobiology Center of NINS, 2-21-1 Osawa, Mitaka, Tokyo
181-8588, Japan}

\author[0000-0002-0746-1980]{Jacob Lustig-Yaeger}
\affiliation{Department of Astronomy and Astrobiology Program, University of Washington, Box 351580, Seattle, Washington 98195, USA}
\affiliation{NASA Astrobiology Institute's Virtual Planetary Laboratory, Box 351580, University of Washington, Seattle, Washington 98195, USA}


\begin{abstract}

In the near future, extremely-large ground-based telescopes may conduct some of the first searches for life beyond the solar system.  High-spectral resolution observations of reflected light from nearby exoplanetary atmospheres could be used to search for the biosignature oxygen. However, while Earth's abundant \ce{O2} is photosynthetic, early ocean loss may also produce high atmospheric \ce{O2} via water vapor photolysis and subsequent hydrogen escape. To explore how to use spectra to discriminate between these two oxygen sources, we generate high-resolution line-by-line synthetic spectra of both a habitable Earth-like, and post-ocean-loss Proxima Centauri b. We examine the strength and profile of four bands of \ce{O2} from 0.63 to 1.27 $\mu$m, and quantify their relative detectability.  We find that 10 bar \ce{O2} post-ocean-loss atmospheres have strong suppression of oxygen bands, and especially the 1.27$\mu$um band. This suppression is due to additional strong, broad \ce{O2}-\ce{O2} collisionally-induced absorption (CIA) generated in these more massive \ce{O2} atmospheres, which is not present for the smaller amounts of oxygen generated by photosynthesis. Consequently, any detection of the 1.27$\mu$m band in reflected light indicates lower Earth-like \ce{O2} levels, which suggests a likely photosynthetic origin. However, the 0.69 $\mu$m O$_2$ band is relatively unaffected by \ce{O2}-\ce{O2} CIA, and the presence of an ocean-loss high-\ce{O2} atmosphere could be inferred via detection of a strong 0.69 $\mu$m O$_2$ band, and a weaker or undetected 1.27 $\mu$m band. These results provide a strategy for observing and interpreting \ce{O2} in exoplanet atmospheres, that could be considered by future ground-based telescopes. 

\end{abstract}

\keywords{planets and satellites: atmospheres -- planets and satellites: individual (Proxima Centauri b) -- planets and satellites: terrestrial planets -- techniques: spectroscopic}

\section{Introduction}

\label{sec:intro}

The search for life beyond the Solar System will enter a new phase over the next decade, with near-term attempts to characterize M dwarf terrestrial exoplanet atmospheres using the James Webb Space Telescope and extremely large telescopes on the ground.  One of the most promising techniques to search for \ce{O2} in exoplanet atmospheres is high-resolution spectroscopy with ground-based telescopes \citep{Snellen2013,Snellen2015,Rodler2014,HawkerPerry2019,Lopez-MoralesMercedes2019DEBo}. In the near future, instruments such as the Near Infrared Spectrograph \citep{Lee2010} and the  G-CLEF spectrograph \citep{Szentgyorgyi2016} on the Great Magellanic Telescope (GMT), HIRES on the Extremely Large Telescope \citep{Marconi2016}, or HROS and NIRES on the Thirty Meter Telescope (TMT) \citep{Froning2006} will enable high-resolution spectroscopy of terrestrial exoplanets. \ce{O2} may be one of the most readily detectable biosignature gases to search for, because it is produced by photosynthesis, and can accumulate to high abundance in planetary atmospheres \citep{Meadows2017a,Meadows2018b}. 

\begin{figure}[bt]
    \centering
    \includegraphics[width=0.47\textwidth]{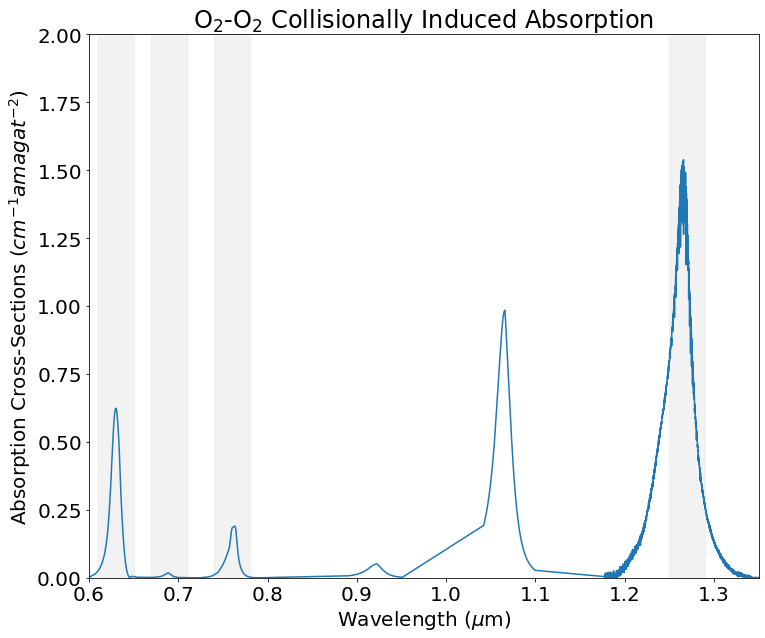}
    \caption{Wavelengths at which \ce{O2}-\ce{O2} collisionally induces absorption \citep{Karman2019} in cross sectional units, versus wavelength. Grey bars indicate O$_2$ bands discussed throughout the paper.}
    \label{fig:cia_plot}
\end{figure} 

\begin{figure*}[tb]
    \centering
    \includegraphics[width=0.97\textwidth]{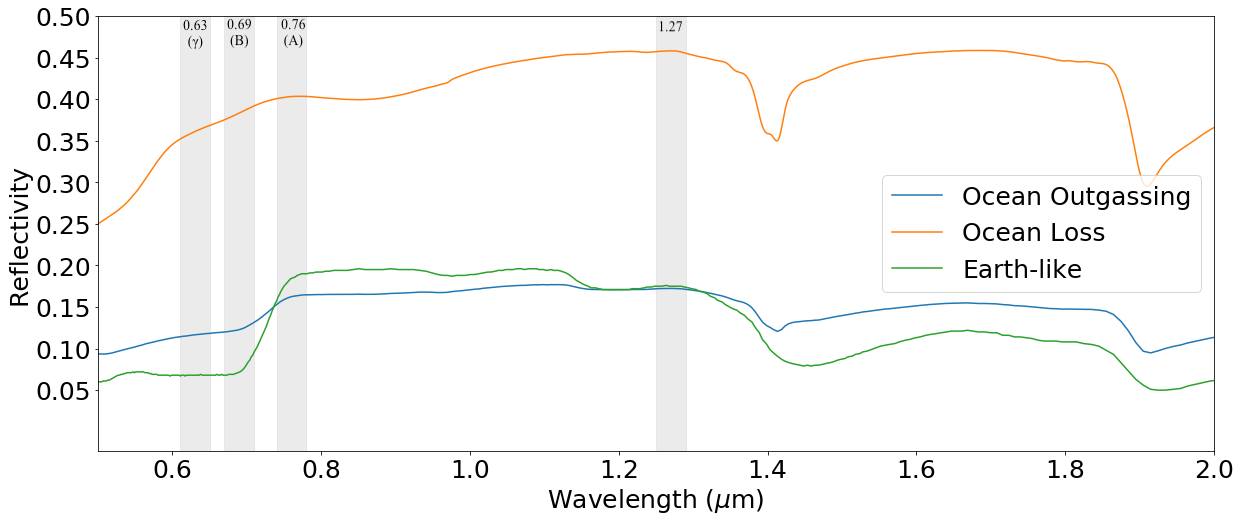}
    \caption{Surface albedos used in three atmospheres with gray marker bars indicating oxygen bands examined later. Each model used a unique albedo representative of potential surface conditions: Ocean Outgassing uses an Earth albedo with vegetation removed, Ocean Loss uses a Desert Albedo, and Earth-like uses an average Earth albedo.}
    \label{fig:albedo}
\end{figure*}

\begin{figure*}[tb]
    \centering
    \includegraphics[width=0.45\textwidth]{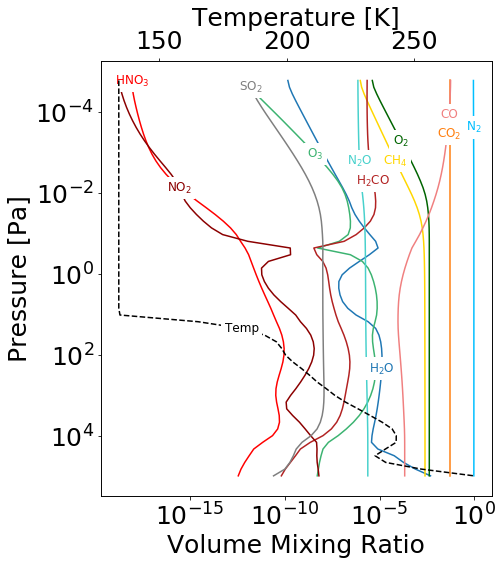}
    \includegraphics[width=0.45\textwidth]{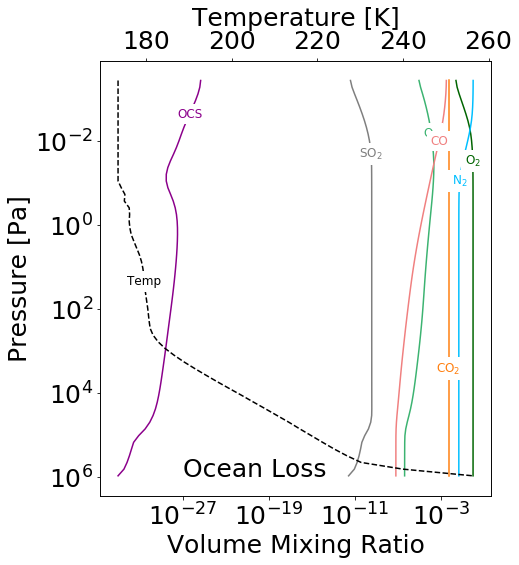}
    \includegraphics[width=0.45\textwidth]{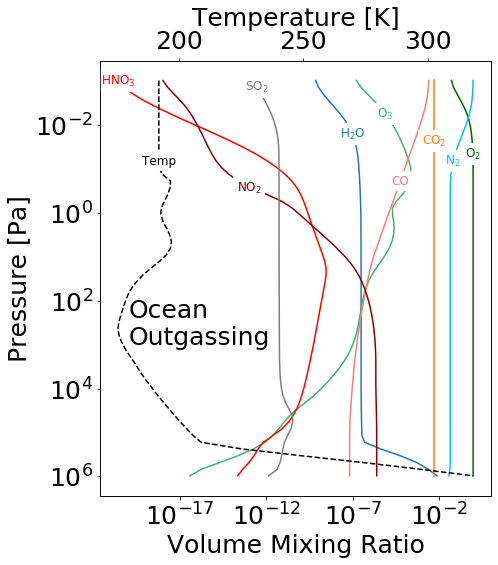}
    \caption{Gas mixing ratios varying with pressure and temperature in the selected atmospheric models of Proxima Centauri b. Top row, from left to right:  1 bar Earth-like,  10 bar post-ocean-loss. Bottom row: 10 bar ocean outgassing}
    \label{fig:ptplots}
\end{figure*}

\begin{figure}[bt]
    \centering
    \includegraphics[width=0.47\textwidth]{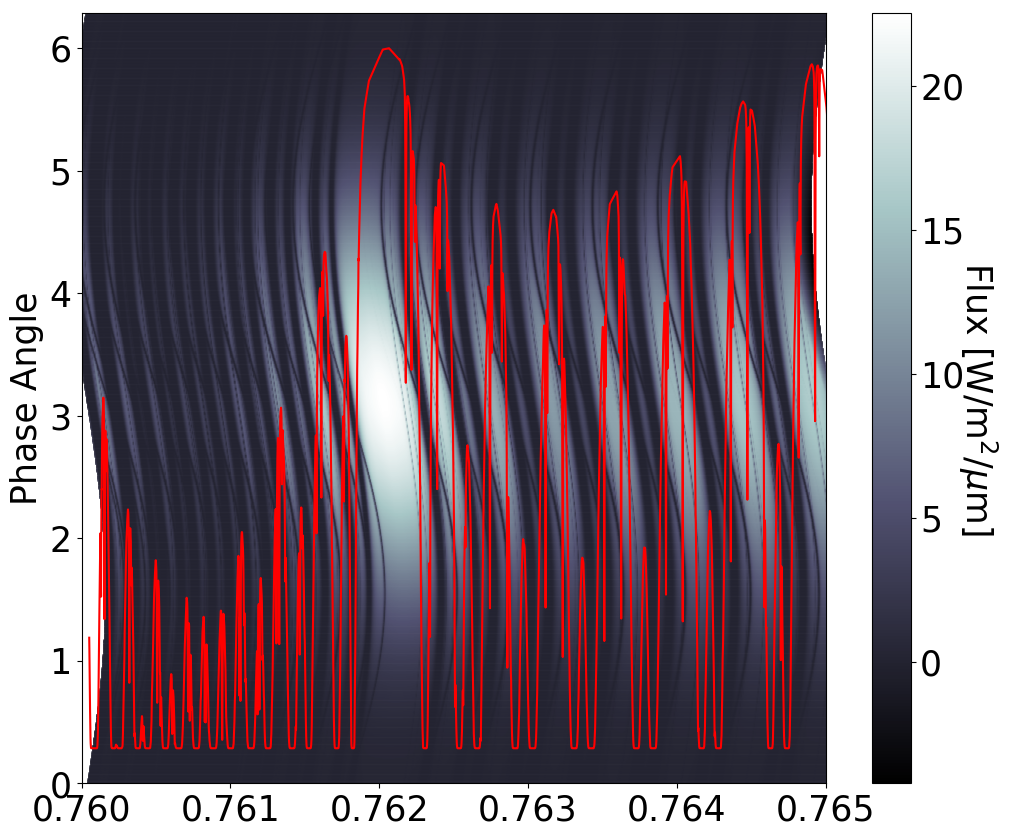}
    \caption{Phase-dependent radial velocity shift of the 0.76 $\mu$m oxygen absorption band. For this calculation, we used accepted values for the barycentric and systemic velocities of the Proxima Centauri system. Our results agree with \cite{Lovis2017} in this determination of radial velocity and are used consistently throughout this paper.Earth's telluric oxygen lines are overlaid in red. This demonstrates the ability to separate observations from telluric lines using high-resolution spectroscopy. Note that observation is optimized for phases close to quadrature. }
    \label{fig:rvplot}
\end{figure}

\begin{figure}[tb]
    \centering
    \includegraphics[width=0.45\textwidth]{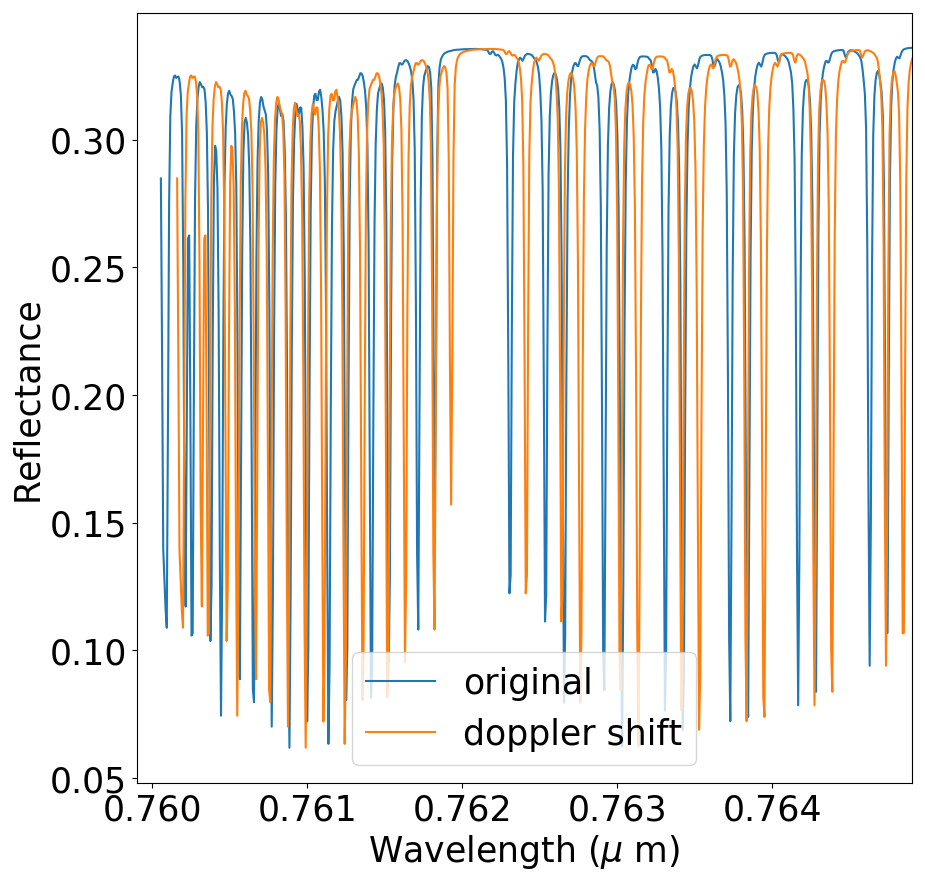}
    \caption{Clear separation shown between telluric and radial velocity shifted lines for Proxima Centauri b. This is one example of the doppler shift due to the orbital, barycentric, and systemic radial velocities. Note that some shifted lines move onto other telluric lines compounding the separation. }
    \label{fig:separation}
\end{figure}

Near-term terrestrial exoplanet characterization will focus on M dwarf systems. This is due in part to the prevalence of M dwarf stars in our galaxy, and the relative ease with which small planets can be detected and characterized when orbiting these diminutive stars. Many of the already detected terrestrial planets orbit M dwarfs \citep{Anglada2016, Gillon2017, Berta2015, Bonfils2018,Dittmann2017}. Based on analysis of Kepler data and considering conservative and optimistic limits for the habitable zone (HZ), it is estimated that there are 0.16-0.24 Earth-sized ($<1.6$R$_\oplus$) and 0.12-0.21 super-Earth sized terrestrial planets per M dwarf HZ \citep{Dressing2015}. This shows a clear opportunity for further study as terrestrial M dwarf planets orbiting in the habitable zone are likely to be abundant and have the potential to be habitable. 

As noted by \citet{Snellen2015} and 
\citet{Lopez-MoralesMercedes2019DEBo}, high resolution ground based spectroscopy is particularly well suited to observing Earth-size planets orbiting M dwarfs due to the favorable planet-to-star flux contrast ratios.  The high-resolution technique can be used to observe planetary atmospheres in transmission, for transiting M dwarf planets, or in reflected light, for planets that do not transit. The latter capability is enabled by the orbital motion of the planet, which allows absorption in reflected light from the planet's atmosphere to be separated from absorption in the Earth's atmosphere along the telescope's line of sight. Being able to study non-transiting planets means that observationally-accessible targets can be identified nearer to our solar system, because the probability of transit is relatively low, meaning that transiting planets are typically much further from us.  The closest known exoplanet, Proxima Centauri b \citep{Anglada2016}, orbits in the habitable zone of its parent star, but  does not transit, and it is an excellent target for future high-resolution spectroscopic study \citep{Lovis2017}.  

While planets orbiting M dwarfs may be more observationally favorable, these planets may be subjected to stellar radiative and gravitational processes that could adversely impact their potential habitability.  Compared to G dwarf stars like our Sun, M dwarfs are highly active, with frequent flares, strong stellar winds and other stellar events that could potentially modify the composition or remove a planetary atmosphere \citep{Segura2010,Erkaev2007,Shields2016b}. Furthermore, due to their proximity to their star, M dwarf planets may be tidally locked and, if in circular orbits, synchronously rotating \citep[e.g][]{Ribas2016}. However, if the planets' orbits are eccentric, they may be subject to tidal heating due to temporally-varying gravitational field gradients, that could even drive ocean loss \citep{Barnes2013a}. 

However, perhaps the most significant known impact on the potential for habitability for close-in M dwarf planets is the luminosity evolution of M dwarf stars.  Early, pre-main-sequence M dwarfs are super-luminous as they contract slowly down to their main sequence size and luminosity \citep{Dotter2008,Baraffe2015}.  Planets that form in what will become the main sequence habitable zone are therefore subjected to very high luminosities early on. This could lead to a runaway green house resulting in the vaporization of any surface water  \citep{Kleine2009, Kopparapu2013a}.  Photolysis of this water in the upper atmosphere, and subsequent H escape to space, could leave behind high amounts (~1000 bar) of atmospheric oxygen \citep{Chassefiere1997,Luger2015a}. 
However, \ce{O2} atmospheric loss processes and surface sinks may reduce this ocean-loss \ce{O2} atmosphere to 10s of bars or less. This oxygen could mimic biologically-generated oxygen from photosynthesis and represent a false positive for life on a post-ocean-loss world \citep{Luger2015a, Schwieterman2016, Lustig-Yaeger2019}.

Consequently it may be that \ce{O2} in M dwarf planetary atmospheres may be a common outcome of their early evolution, even for planets in what becomes the main-sequence habitable zone \citep{Luger2015a,Lincowski2018}, and the challenge will be to determine whether the \ce{O2} detected is due to ocean loss, or a true photosynthetic biosphere. Discrimination between whether a biotic \citep{Gebauer2018} or abiotic source is responsible for an M dwarf planet's atmospheric oxygen may be possible through examination of other environmental features, including other gases in the atmosphere \citep{Domagal-Goldman2014,Meadows2017a}. These false-positive discriminants have been studied intensively for a variety of different false positive mechanisms for low-resolution spectra.  In the ocean loss case, the high-\ce{O2} atmosphere that remains may be of sufficiently high pressure, potentially several bars or more \citep{Luger2015a, Schaefer2016}, that it produces \ce{O2}-\ce{O2} collisionally-induced absorption.  This absorption can be quite strong and broad, and for low resolution spectra, models predict that \ce{O2}-\ce{O2} may be detectable with the upcoming James Webb Space Telescope, and would be a strong indicator of past ocean loss \citep{Schwieterman2016,Lincowski2018,Lustig-Yaeger2019}.

However, near-term observations of \ce{O2} on extrasolar planets in the habitable zone may also be possible from the ground.   Specifically for the closest known exoplanet, Proxima Centauri b, using a combination of high-contrast imaging and high-resolution spectroscopic observations in reflected light \citep{Lovis2017}. Although terrestrial \ce{O2} may be feasible to observe in the near-term with high-resolution spectroscopy, very little work has been done on determining techniques to differentiate between atmospheric \ce{O2} generated by a photosynthetic biosphere, and \ce{O2} due to ocean loss processes.  Here we present simulated high-resolution spectra for Proxima Centauri b as a self-consistent Earth-like planet with a photosynthetic biosphere, and as a planet with a high-\ce{O2} atmosphere as the result of ocean loss. We use these spectra to explore the feasibility of identifying ocean loss for M dwarf exoplanets, and to develop recommendations for optimum observing techniques to do so.  

\section{Methods}
\label{sec:methods}

\begin{figure*}[bt]
    \centering
        \includegraphics[width=0.95\textwidth]{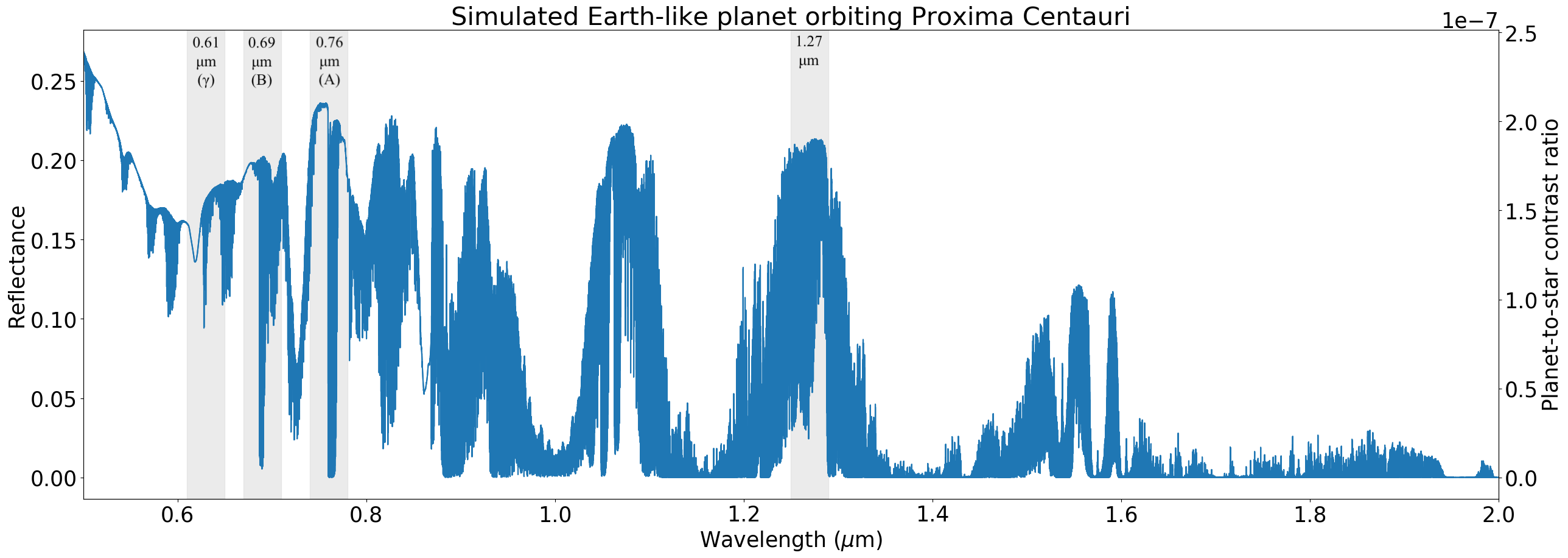}
        \includegraphics[width=0.95\textwidth]{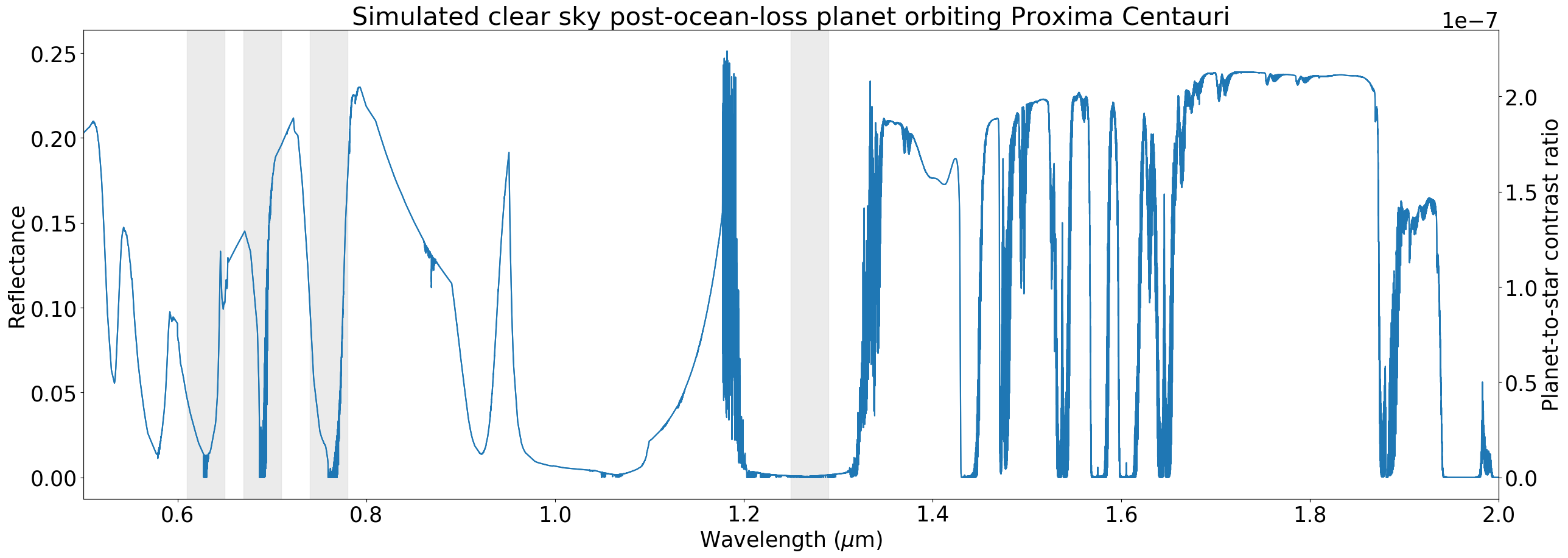}
        \includegraphics[width=0.95\textwidth]{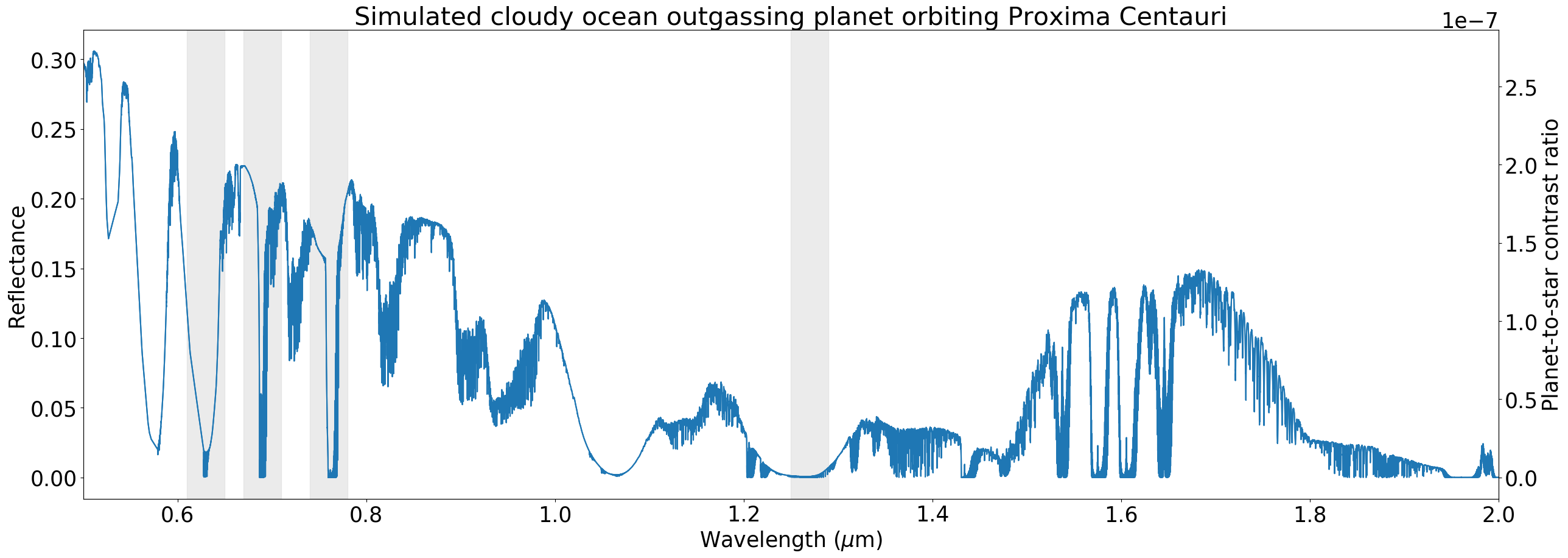}
    \caption{Comparison of a photochemically self-consistent Earth-like atmosphere with two different 10 bar high-oxygen, ocean-loss atmospheric models. The vertical grey bars show oxygen bands centered at the labeled wavelengths. All planetary parameters besides the atmospheric composition and albedo are held constant, so this plot demonstrates self consistent spectral differences due to planetary conditions and it provides broad-wavelength context for later figures shown at much higher resolution. Significant differences can be seen between the Earth-like spectrum and the ocean-loss worlds in the relative width of the molecular absorption bands. Note that these spectra are shown here in full phase for illustrative purposes to highlight absorption.}
    \label{fig:context}
    \end{figure*}
    
    \begin{figure*}[bt]
    \centering
     \includegraphics[width=0.35\textwidth]{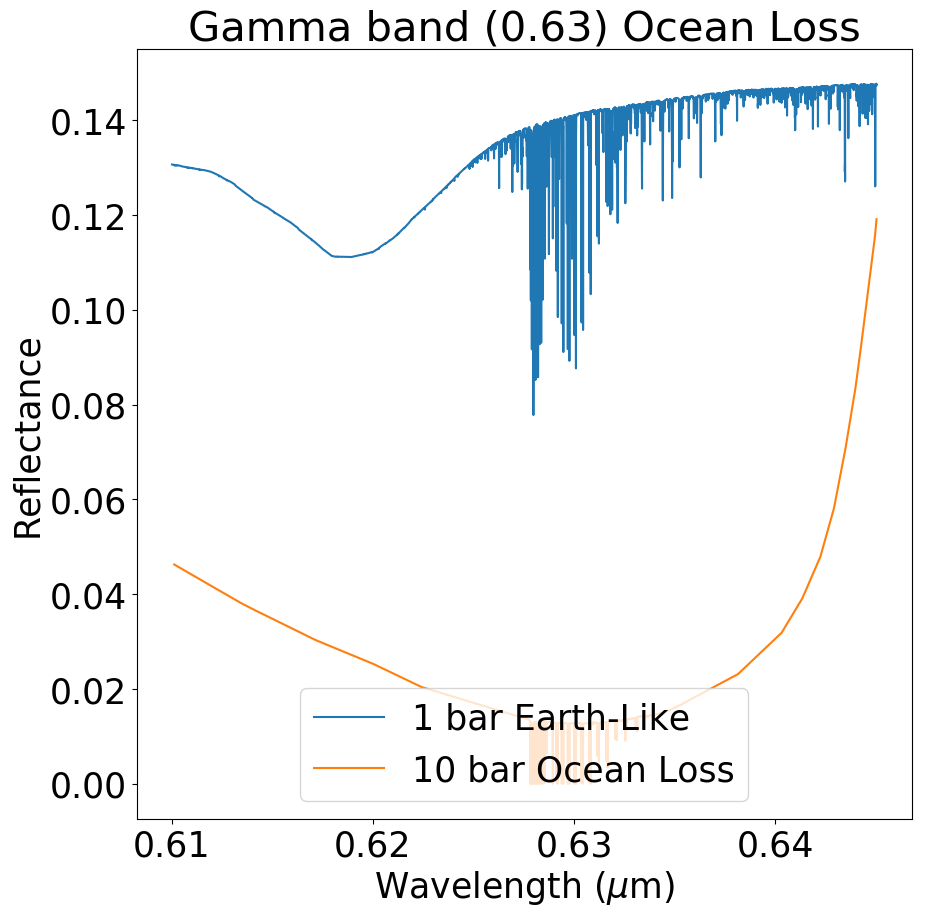}
    \includegraphics[width=0.35\textwidth]{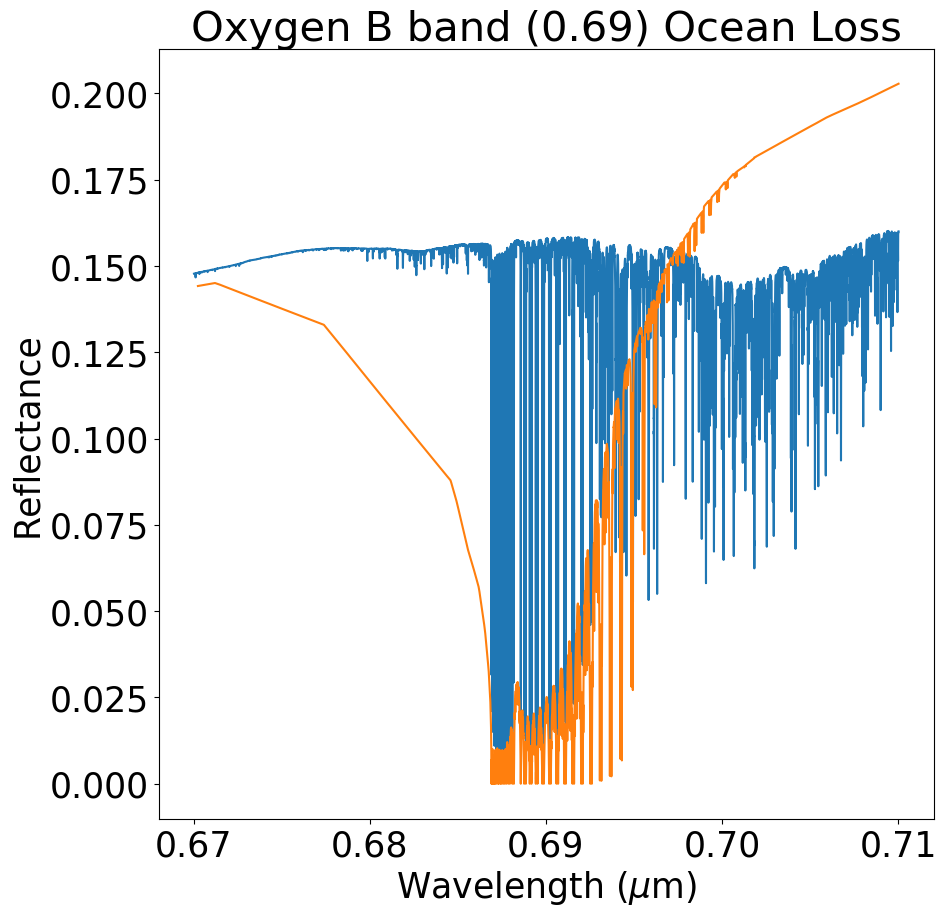}'
    \includegraphics[width=0.35\textwidth]{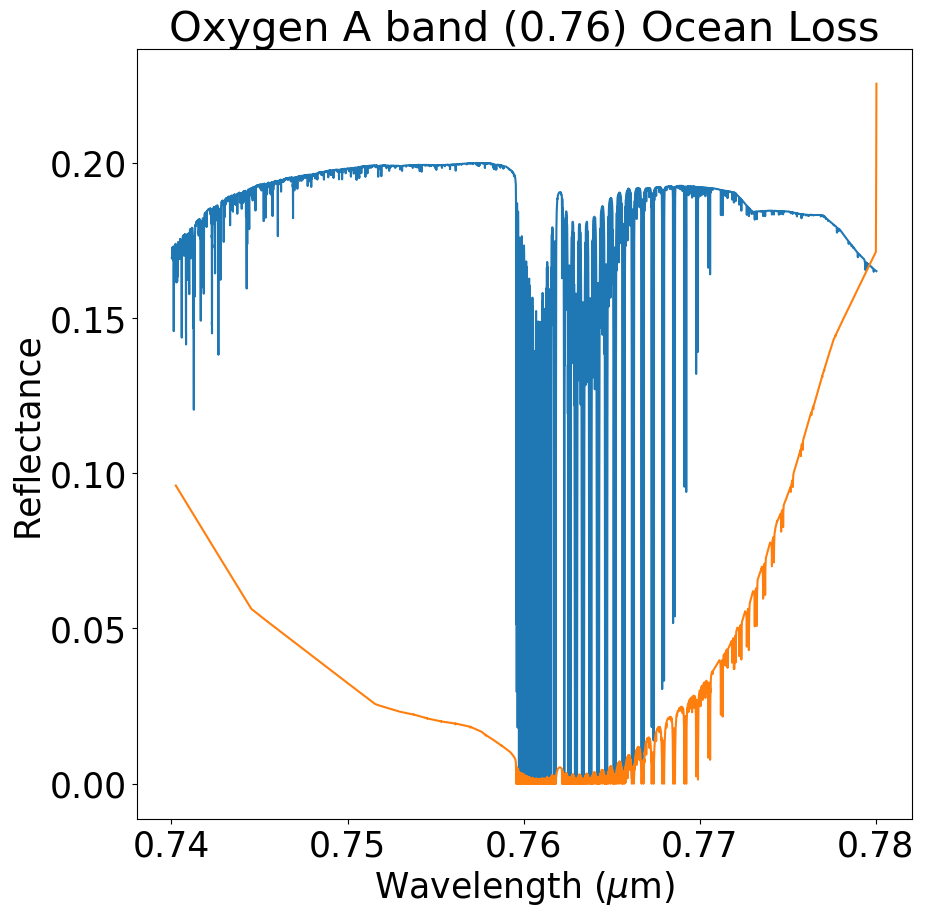}
    \includegraphics[width=0.35\textwidth]{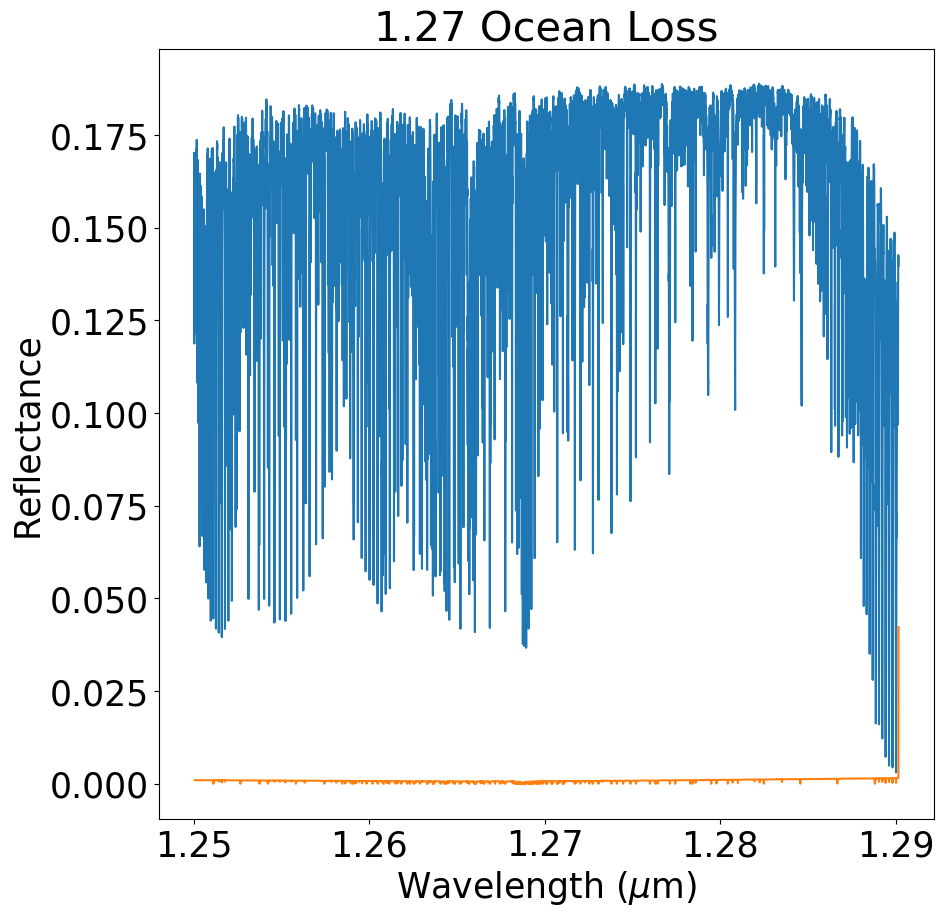}

     \caption{Comparison of spectra for a self-consistent Earth-like Proxima Centauri b atmosphere (blue) and an ocean loss 10 bar O$_2$ atmosphere (orange). Demonstrates the suppression of the oxygen absorption in the ocean loss case due to strong \ce{O2}-\ce{O2} collisionally-induced absorption bands}
    \label{fig:ol}
    \end{figure*}

In this paper, we used a newly-developed model for high-resolution terrestrial exoplanet spectra, that can simulate radial velocity shifts, if needed, and uses a sophisticated, line-by-line atmospheric radiative transfer model to simulate the high-resolution spectra of Proxima Centauri b for two primary types of atmospheres: high oxygen (post-ocean-loss) and self-consistent Earth-like, where the composition of the atmosphere is determined from prescribed surface fluxes and photochemistry driven by the spectrum of Proxima Centauri b. 

\subsection{SMART Radiative Transfer Model} 

To generate these high resolution spectra, we used the line-by-line Spectral Mapping and Atmospheric Radiative Transfer code \citep[SMART; originally developed by D. Crisp][]{Meadows1996}. SMART was specifically developed to model solar system terrestrials and has been validated against several, including Earth and Venus \citep{Tinetti2005, Robinson2011, Arney2014}. SMART is uniquely well suited for high-resolution spectral simulation as it has a high native spectral resolution, which is typically degraded to lower resolution for other practical applications. Additionally, SMART is highly flexible to user need with the ability to choose wavelength ranges and resolution as well as incorporating inputs including collision-induced absorption data, spectral surface albedo data, and vertically resolved gas mixing ratios. SMART also uses Rayleigh coefficents from \cite{Young1980}. For the high oxygen cases, we used \ce{O2} Rayleigh scattering instead of air which is used for the Earth-like case.

This code can also model aerosols such as hazes or clouds to predict their impact on the spectrum of an atmosphere. In this paper, we simulate an Earth-like atmosphere with 50\% clear sky and 50\% of an equal mixture of cirrus and stratocumulus clouds which are assumed to be uniformly-spaced around the globe. Because our models are 1-D and produce globally-averaged environments and spectra they may not accurately represent the cloud distribution on synchronously-rotating  planets, which 3D GCM models of 1-bar atmospheres have suggested may be concentrated near the subsolar point on these slowly-rotating planets \citep{DelGenio2019Proxima,Yang2017}. In this scenario, cloud reflectivity from the asymmetric cloud-distribution would have a strong phase-dependent effect, and the clouds may form at higher altitudes. However, this is unlikely to impact the main conclusions of our work, as observations of the planet would be taken at close to the same phase, and the strength of the \ce{O2}-\ce{O2} CIA is still significantly stronger in the 10 bar atmospheres when compared to the Earth-like 1 bar atmosphere, even in the presence of 100\% cloud cover, because the clouds are more likely to condense near the 4-5 bar level.

\subsection{Model Inputs} \label{sec:methods:inputs}

To generate the absorption coefficients needed by SMART, we use the Line-By-Line Absorption Coefficients code \citep[LBLABC;][]{Meadows1996}. LBLABC uses a specified atmosphere file containing temperature, pressure and vertical distribution of atmospheric gases. This information is combined with the molecular line lists provided in HITRAN 2016 \citep{Gordon2017} to calculate rotational-vibrational line absorption coefficients to use in the radiative transfer calculations. We used the most recent HITRAN \ce{O2}-\ce{O2} Collisionally Induced Absorption as updated in \cite{Karman2019}. The \ce{O2}-\ce{O2}  relative strengths we used can be seen in Figure \ref{fig:cia_plot} . 

In this paper, we examine three different modelled atmospheres for Proxima Centauri b (PCb) from \cite{Meadows2018a}.  The first two simulate different possible outcomes for a post-ocean-loss abiotic oxygen atmosphere. One such scenario is losing the entire ocean to space. This results in a completely desiccated planet with no water vapor and high levels of atmospheric oxygen. This model has no aerosols present in the atmosphere. The other option is that the planet experiences significant ocean loss and oxygen buildup, but does not lose its entire water inventory. In this case the planet can outgas additional volatiles from its interior such that it has an ocean and atmospheric water vapor, along with an oxygen-dominated atmosphere. Both of these models assume 10 bars of oxygen. The oxygen outgassing simulation has stratocumulus clouds at 5.4 bars and cirrus clouds at 2.4 bars. The atmosphere is weighted 50\% clear sky, and 25\% of each of the cloud types. The third model atmosphere is an Earth-like model, primarily shown in this paper for comparison as a standard for biogenic oxygen. This model uses an Earth-like atmosphere from \cite{Meadows2018a} that is photochemically self-consistent with the M dwarf stellar SED and includes 50\% cloud cover as described in \ref{sec:methods:inputs}.  Pressure-temperature profiles  and abundances for these three models can be seen in Figure \ref{fig:ptplots}.

These three climate scenarios have different surface albedo inputs meant to consistently reflect the assumed surface conditions. The desiccated oxygen world has a desert albedo, the ocean oxygen planet has a modified Earth albedo with vegetation removed, and the Earth-like scenario uses a composite ocean albedo that reproduces the modern Earth surface spectrum. These input surface albedo spectra are shown in Figure \ref{fig:albedo}; see \citet{Meadows2018a} for more details on their composition.

The stellar spectrum used is the Proxima Centauri spectrum from \cite{Meadows2018a}. This spectrum was created by combining available data and modelling wavelengths above 30 $\mu$m as a black body. For the wavelength range examined in this paper, data from the HST Faint Object Spectrograph (up to 0.850 $\mu$m) was combined with calculations using the PHOENIX spectral library v2.0 models in the infrared to generate the stellar spectra. The full stellar spectrum and comparison to the solar spectrum can be seen in Figure 2 of \cite{Meadows2018a} and is available for download on VPL website\footnote{\url{http://vpl.astro.washington.edu/spectra/stellar/proxcen.htm}}. Note that the Proxima spectrum used here is of lower spectral resolution (R $\approx$ 2000-5000) than the resulting planetary simulations (R=100,000). However, this is unlikely to impact our results as we do not calculate noise in these simulations.

\subsection{Phase-Dependent Effects}

We consider both the radial velocity Doppler shift of the planet's reflected spectrum and the phase-dependent brightness changes as a function of orbital phase. 
Our calculation for the Doppler shift combines the systemic radial velocity of Proxima Centauri, the orbital motion of Proxima Centauri b, and the shift from Earth's orbit, and is based on the calculations in \cite{Lovis2017}. The radial velocity shift enables ground based spectrometers to separate telluric lines from observations as shown in Figure \ref{fig:separation}. 
We also calculate how the planet's brightness changes as a function of orbital phase by assuming it scatters light as a Lambertian sphere \citep{Robinson2016}. 

Figure \ref{fig:rvplot} shows both Doppler shifted spectral lines and the Lambertian brightness changes using color contours of planet flux as a function of wavelength and orbital phase. The radial velocity shift can be distinguished relative to the Earth's telluric lines for the selected spectral bands. The planet peaks in brightness at an orbital phase of $\pi$, when there is no orbital radial velocity shift to the planet's spectrum. We combine these effects into a detectability metric in the next section. 

\begin{figure*}[bt]
    \centering
    \includegraphics[width=0.35\textwidth]{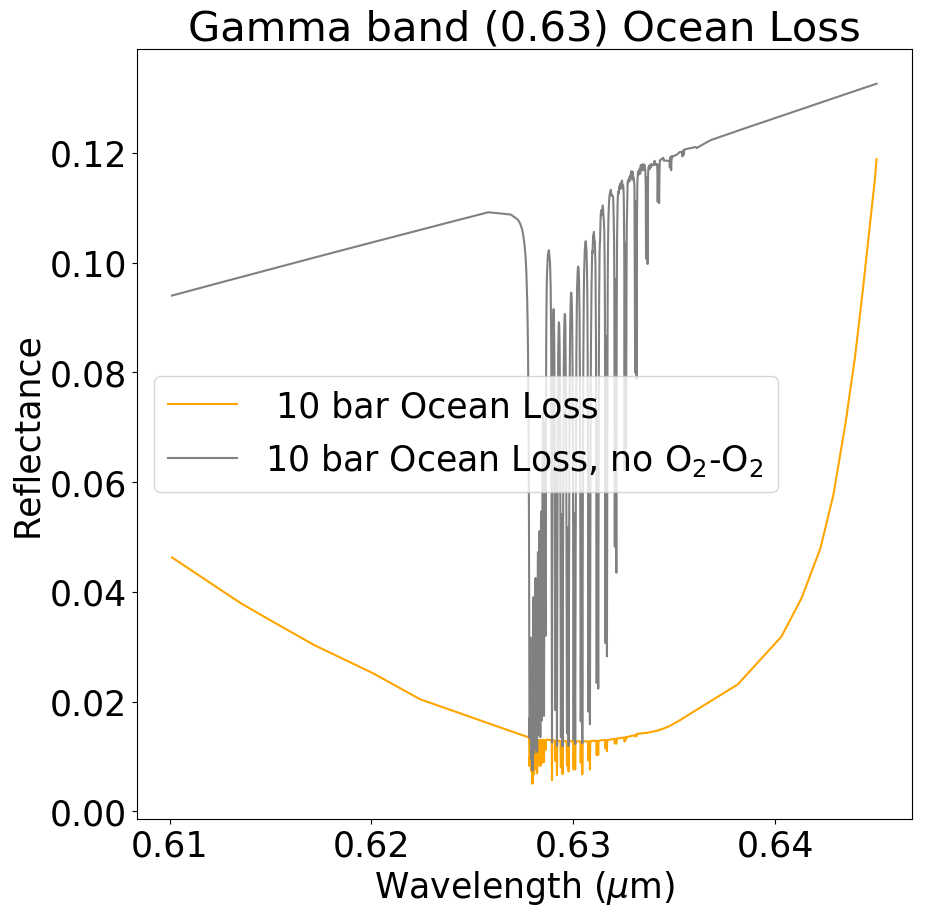}
    \includegraphics[width=0.35\textwidth]{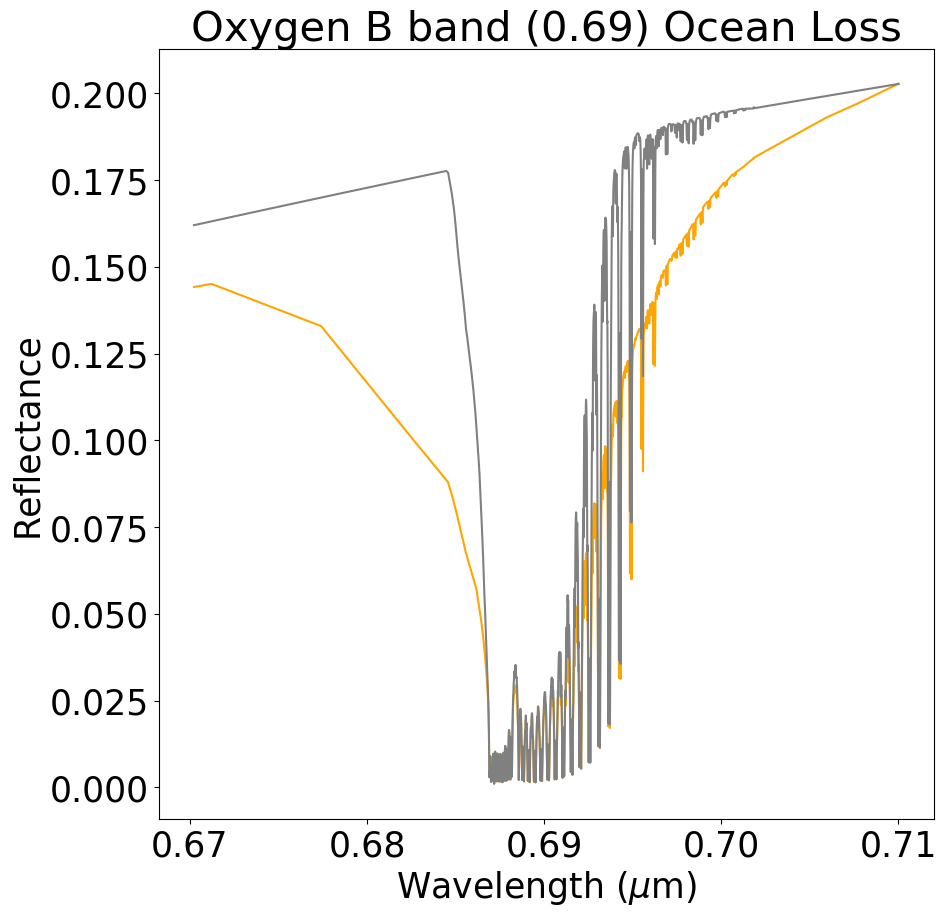}
    \includegraphics[width=0.35\textwidth]{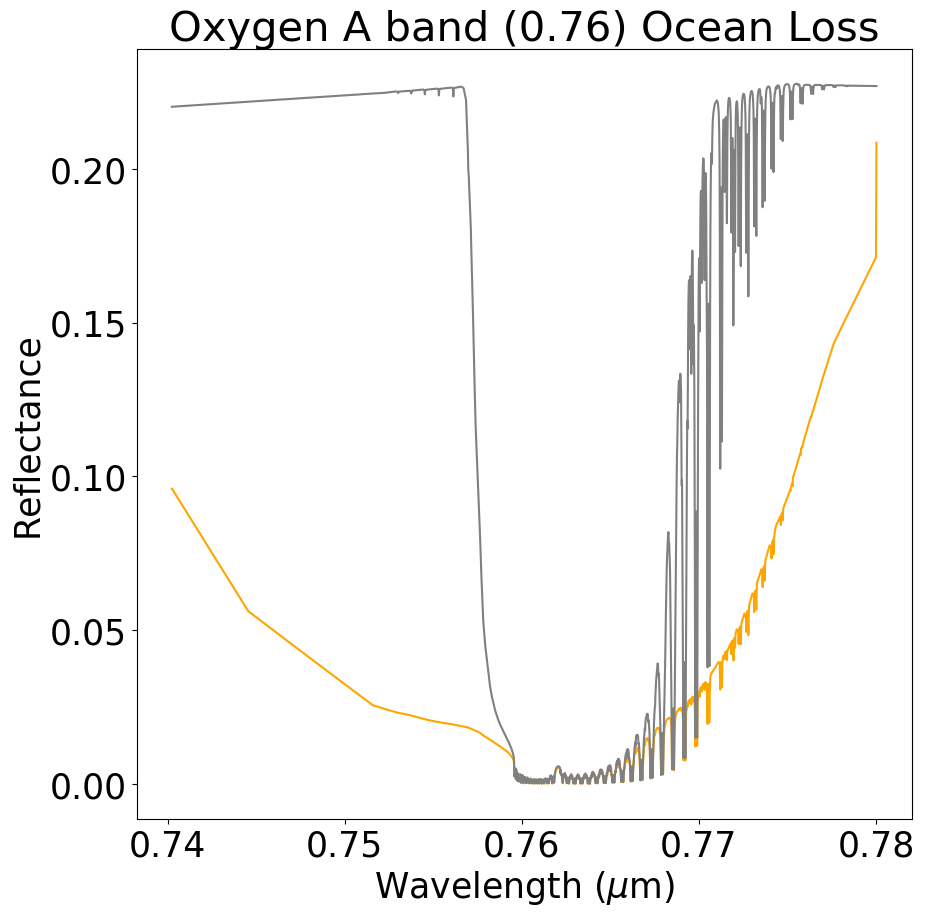}
     \includegraphics[width=0.35\textwidth]{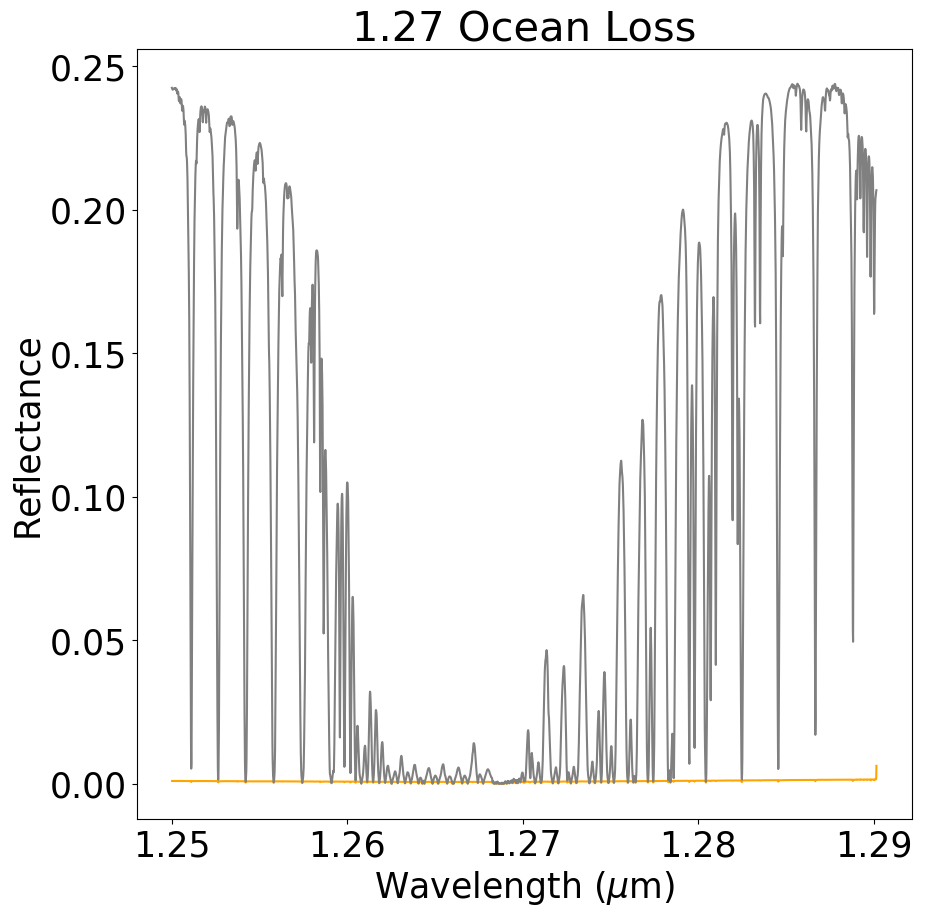}
    \caption{Comparison of 10 bar ocean loss scenario with (orange) and without (grey) the effect of O$_2$-O$_2$ collisions. Demonstrates the effect of the CIA in suppressing the oxygen absorption bands, especially at 1.27 $\mu$m.}
    \label{fig:ol_noO4}
\end{figure*}
    
\begin{figure*}[bt]
    \centering
     \includegraphics[width=0.35\textwidth]{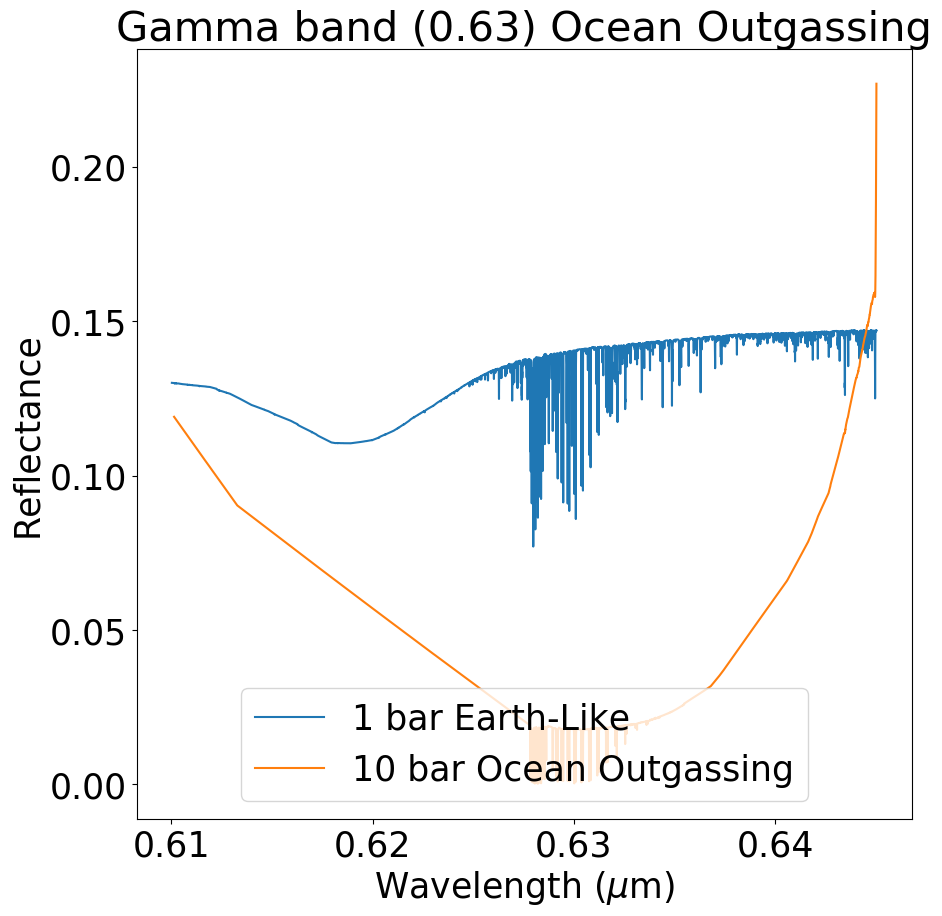}
     \includegraphics[width=0.35\textwidth]{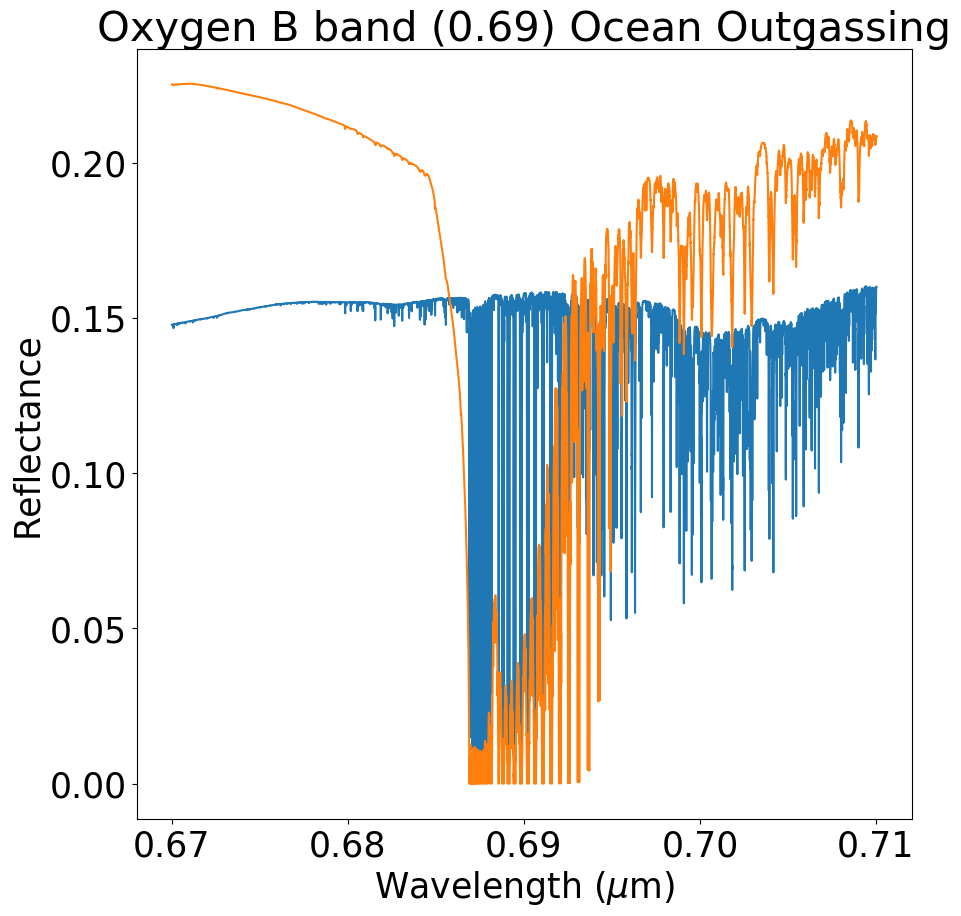}
    \includegraphics[width=0.35\textwidth]{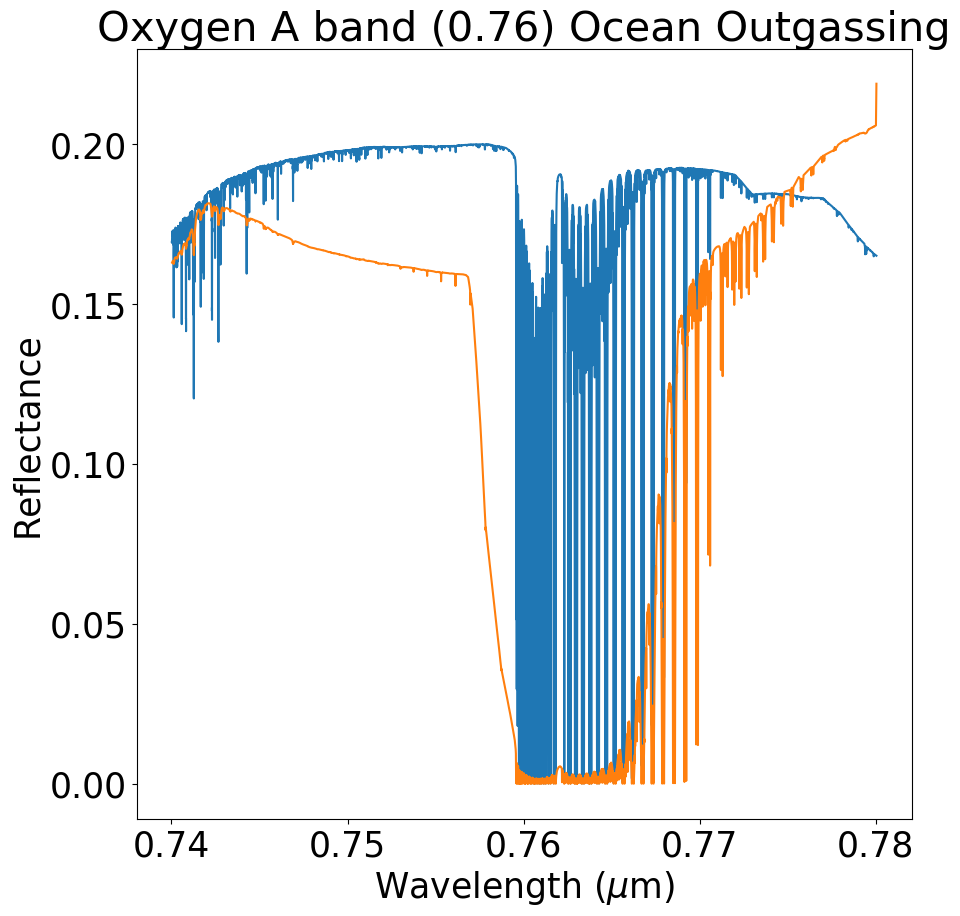}
    \includegraphics[width=0.35\textwidth]{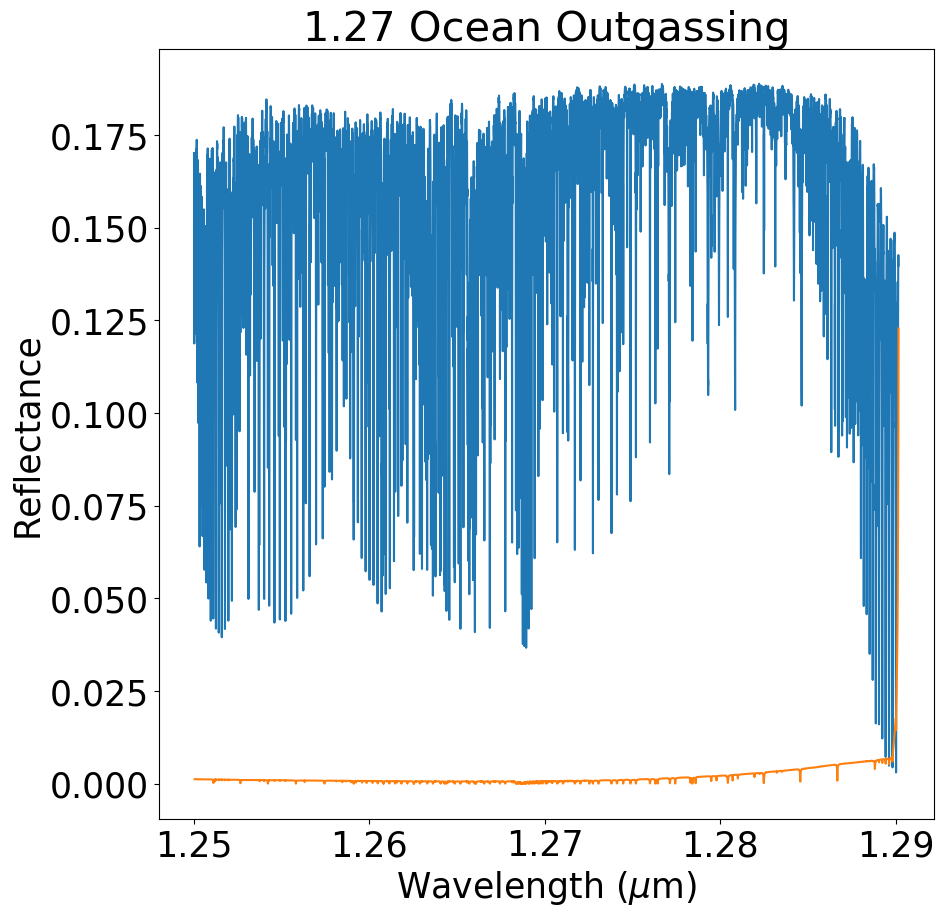}
    \caption{Highlighted oxygen bands in an ocean outgassing 10 bar O$_2$ atmosphere. This atmosphere also includes water vapor and other volatiles in limited quantities.}
    \label{fig:oo}
\end{figure*} 

\subsection{Additional Parameters and Detectability Calculations}

Our spectra are simulated with a spectral resolution $R = 100,000$ to best match the near-future capabilities for ground based high-resolution spectrometers. They are modelled for Proxima Centauri b using the best estimates for the orbital and planetary radii; 0.0485 AU and 6850 km respectively \citep{Anglada2016}. 

Calculations, comparisons, and plots created use the planet reflectivity or reflectance, $ \pi I / F $, where $I$ is the outgoing top of atmosphere radiance and $F$ is the stellar flux incident at the top of the atmosphere. This is used instead of only planetary flux to remove the spectral contribution of the stellar flux from the planet flux, and to focus exclusively on features in the planet spectrum \citep{Meadows2018a}. 

To quantify the detectability of individual spectral bands we use an integration scheme similar to the methods used to extract exoplanet spectra at high resolution. This method is based off of that used in \cite{Snellen2017}.
We define the band-integrated detectability, 
\begin{equation}
    \mathcal{D}_{\lambda} = \int_{\lambda_0}^{\lambda_1} | F_p(\lambda) - \bar{F}_p(\lambda) | d\lambda
    \label{eqn:detect1}
\end{equation}
where $F_p(\lambda)$ is the high resolution planet spectrum, $\bar{F}_p(\lambda)$ is the low-resolution planet spectrum, and the integral extends from the molecular band minimum ($\lambda_0$) to the band maximum ($\lambda_1$). In order to get these varying spectra, we ran SMART at different resolutions. This formula isolates the absorption features as deviations from the continuum.
This approach captures both the magnitude and frequency of the lines making up a band to obtain a quantitative proxy for detectability, and allows multiple bands to be compared. 

We also use a similar approach to quantify the detectability of individual molecular bands after taking into account Earth's telluric absorption and the phase-dependent radial velocity Doppler shift of the planet spectrum. We define the phase-integrated detectability, 

\begin{equation}
    \mathcal{D}_{\lambda, \alpha} = \int_{0}^{2\pi} \int_{\lambda_0}^{\lambda_1}  \mathcal{T}_{\oplus}(\lambda) | F_p(\lambda, \alpha) - \bar{F}_p(\lambda, \alpha) | d\lambda d\alpha
    \label{eqn:detect2}
\end{equation}

where $F_p(\lambda, \alpha)$ is the high spectral resolution phase-dependent planet flux as shown in Figure \ref{fig:rvplot}, which takes into account both the planet's RV shift and overall flux variation, $\bar{F}_p(\lambda, \alpha)$ is the low resolution planet spectrum, and $\mathcal{T}_{\oplus}(\lambda)$ is the Earth's telluric transmittance at the surface due to atmospheric extinction. This phase-integrated detectability metric tends towards large values when the observed planet spectrum is Doppler shifted sufficiently outside the telluric lines. This calculation incorporates all of the phase shifts, and may not accurately reflect an actual observing session that only observes part of the orbit.

\section{Results}
\label{sec:results}

In Figure \ref{fig:context}, we provide plots of the three atmospheres discussed in Section \ref{sec:methods:inputs} to provide broader wavelength context for later discussions. 
The grey bars highlight oxygen bands centered at 0.63 ($\gamma$), 0.69 (B), 0.76 (A), and 1.27 $\mu$m which are examined in depth later. The 10 bar atmospheres show additional absorption and differences in continuum throughout. Due to the updated CIA files used, there are some changes in the Earth-like planet with some features differing from those originally presented in \citet{Meadows2018a}

\subsection{Initial Band Comparison: Ocean Loss}
    
In Figure \ref{fig:ol}, we  examine the strength of the four oxygen bands for the Earth-like model compared with the desiccated high oxygen ocean loss scenario. 

At the oxygen A band, we see pressure broadened wings in the high oxygen atmosphere. The continuum is also decreased and the overall depth and frequency of the lines is strongly diminished. This is due to O$_2$-O$_2$ collisionally induced absorption (CIA) suppressing the spectrum. The continuum has shifted in response, moving from roughly even at 0.25 to a sharp curve from 0. down to less than 0.05.

The oxygen B band centered at 0.69 $\mu$m experiences less additional absorption than the A band. While the ocean loss atmosphere is made up of fewer, shallower lines, they are still relatively similar in magnitude. This band displays a change in shape, but not as dramatic of a reduction in magnitude as seen in other bands. 

For the band centered at 1.27 $\mu$m, the high O$_2$ atmosphere has greatly reduced reflectance, again due to \ce{O2}-\ce{O2} CIA, to the degree that it appears nearly as a flat line with barely perceptible \ce{O2} absorption features on the bottom side. This is in sharp contrast to the deep, frequent lines that make up the 1.27 $\mu$m \ce{O2} band of the Earth-like atmosphere. 

Likewise, for the $\gamma$ band at 0.63 $\mu$m, the high oxygen atmosphere band is shallower; however to a lower degree. In this case, the approximate shape of the band is preserved but overall flux is suppressed. This change is primarily a continuum shift as a result of the different surface albedos of the two simulations. 

\subsection{O$_2$-O$_2$ CIA impact on Ocean Loss planet}

Figure \ref{fig:ol_noO4} highlights the powerful effects of O$_2$-O$_2$ collisionally induced absorption mentioned above. These plots highlight the effect of the O$_2$-O$_2$ CIA in the simulated ocean loss atmosphere by showing spectrum with and without the O$_2$ CIA. It is apparent from this comparison that the O$_2$-O$_2$ CIA is overpowering the structure of the oxygen absorption, effectively lowering the continuum, and leaving the band much shallower. This is particularly the case for the gamma band and the 1.27 $\mu$m band. 
The oxygen A band shows an elevated continuum without the O$_2$-O$_2$ CIA, but maintains the same general structure. The oxygen B band appears less affected by CIA. It is clear from these plots that the O$_2$-O$_2$ collisions have a large effect in an ocean loss atmosphere atmosphere, primarily at 0.63 and 1.27 $\mu$m.

\subsection{Initial Band Comparison: Ocean outgassing planet}

In Figure \ref{fig:oo}, we show the same four oxygen bands simulated on an ocean world. This planet is similar to the previous one examined but has an ocean albedo and outgassing, which provides additional molecular absorption, such as from water vapor. For spectral context, see Figure \ref{fig:context}. 

The oxygen A and B bands have similar shapes to the desiccated spectrum. The continuum is elevated in these plots because they are located in visible wavelengths and are heavily affected by the assumed surface albedo.  

The 0.63 $\mu$m $\gamma$ ocean outgassing band appears similar to the ocean loss case. There is an additional small water vapor feature on the right side but appears mostly the same. The previous details seen in the ocean loss planet comparison hold true in this plot: there is O$_2$-O$_2$ suppression accounting for diminished magnitude in the ocean outgassing scenario. 

At the 1.27 $\mu$m band, the same large degree of supression by \ce{O2}-\ce{O2} is shown with minor differences in the exact reflectance due to other atmospheric gases such as HF, HCl and H$_2$O present. 

\subsection{Oxygen Band Detectability}

\begin{figure*}[bt]
    \centering
    \includegraphics[width=0.9\textwidth]{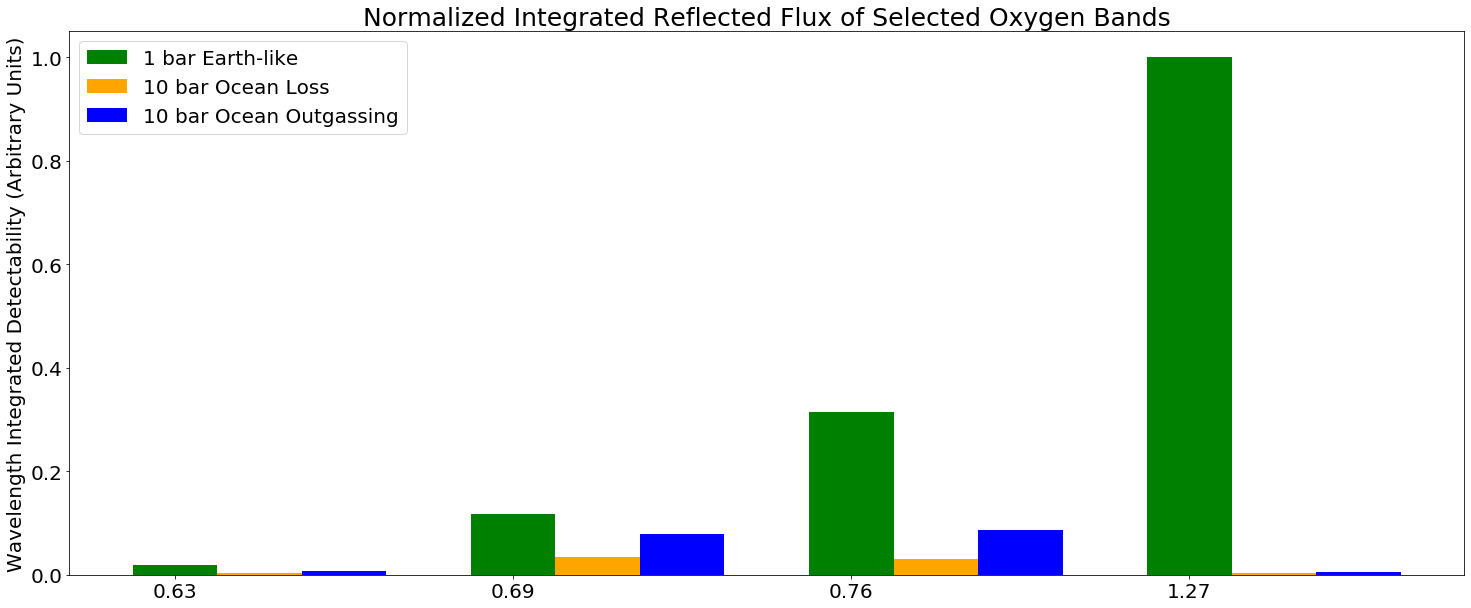}
    \caption{Relative detectability of individual oxygen lines in each band using Equation \ref{eqn:detect1}. Green represents an Earth-like atmosphere, orange a post-ocean-loss atmosphere and blue an ocean outgassing atmosphere. This plot highlights the amount of suppression caused by \ce{O2}-\ce{O2} collisions at the 1.27 $\mu$m band. 
}
    \label{fig:barplot}
\end{figure*} 

\begin{figure*}[bt]
    \centering
    \includegraphics[width=0.9\textwidth]{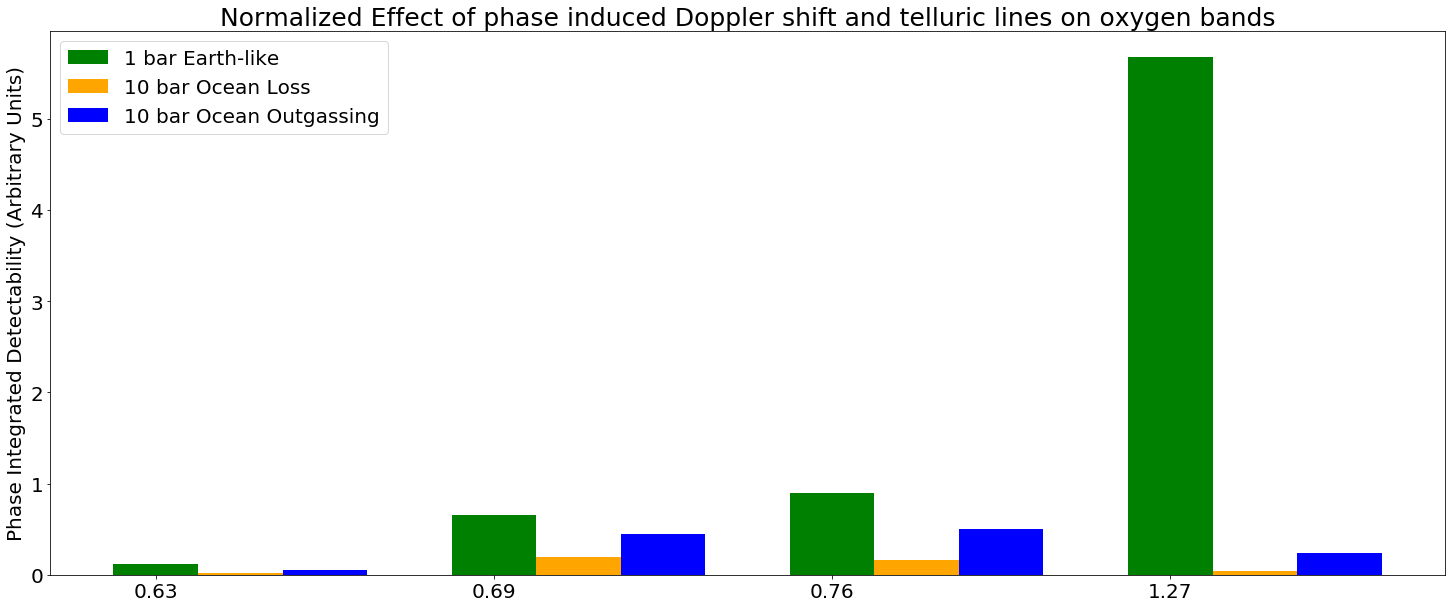}
    \caption{Relative phase-dependent detectability using Equation \ref{eqn:detect2} which incorporates Earth telluric lines and the radial velocity shifted lines. Green represents an Earth-like atmosphere, orange a post-ocean-loss atmosphere and blue an ocean outgassing atmosphere.  This plot highlights that the phase-induced radial velocity shift between telluric and planetary lines produces a negligible effect on detectability compared to the impact of the much broader o2-o2 suppression.  
}
    \label{fig:barplot_shifted}
\end{figure*} 

A quantitative comparison of the relative detectability in the examined bands can be seen in Figures \ref{fig:barplot} and \ref{fig:barplot_shifted}. We used integrated detectability metrics in this case as a proxy for band detectability with the high-resolution cross-correlation technique. See Section \ref{sec:methods} for details about these calculations.  

Figure \ref{fig:barplot} shows a bar plot of the wavelength-integrated detectability for each oxygen band and for each atmospheric case considered. The data are presented in clusters of atmospheres by \ce{O2} band. The green bars represent the Earth-like atmosphere, the orange the ocean loss, and the blue bars are the ocean outgassing planet. The 1.27 $\mu$m has the highest relative detectability of the four bands for the Earth-like case, followed by the 0.76 $\mu$m band, and further decreases with decreasing wavelength.  However, the detectability of the 1.27 $\mu$m band is suppressed by a factor of ${\sim}200$ for both of the high oxygen atmosphere cases compared to the Earth-like case, due to the stronger \ce{O2}-\ce{O2} CIA in this band. In the A, B and $\gamma$ bands, the suppression is present, but to a smaller degree (2-10$\times$ suppressed), with the B band being the least affected by the high oxygen atmosphere. 

Figure \ref{fig:barplot_shifted} is similar to Figure \ref{fig:barplot} and shows the phase-integrated detectability of each oxygen band, for each atmospheric case. In this case, we consider how the planet spectrum Doppler shifts in and out of the telluric band, and how the planet's brightness change with phase may affect the ability to detect the planet's spectrum relative to the Earth's telluric lines. To calculate these values, we use a barycentric velocity value of 29.8 km/s and a systemic velocity of -21.7 km/s. We also incorporated the radial component of Proxima Centauri's orbital velocity. For the barycentric value we assumed the smallest radial velocity shift between Earth and the Proxima Centauri system, which conservatively results in the smallest radial velocity shift between telluric and planetary lines in the spectra. This value also corresponds to that used for previous simulations of high-resolution observations of Proxima Centauri b \citep{Lovis2017}.Note that the phases close to full and zero phase will not be able to be observed but are incorporated into our calculations here since we are using relative values between types of atmospheres.  We find qualitatively and quantitatively similar results to our wavelength-integrated spectral filtering detectability metric, indicating that Proxima Cen b's Doppler shift is sufficient to avoid Earth's telluric oxygen lines for all four bands considered. 

\section{Discussion}
\label{sec:discussion}

Our simulations were performed using a new high-resolution spectroscopic modeling capability which we initially applied to an exploration of how best to discriminate between the source of \ce{O2} in an exoplanet atmosphere, and identify false positives  for \ce{O2} using high resolution spectroscopy.  We found that discriminating the more Earth-like atmospheres that might be produced by a photosynthetic biosphere from ocean loss scenarios with remnant high-\ce{O2} atmospheres may be possible by examining the relative strength of the 1.27 $\mu$m and 0.69 $\mu$m (or 0.79 $\mu$m) oxygen bands. In our simulations, the high-\ce{O2} post-ocean-loss atmospheres tended to have weaker \ce{O2} detectability than the Earth-like case at all wavelengths due to strong O$_2$-O$_2$ CIA over a similar wavelength range to the \ce{O2} bands.  This effect is most pronounced in the 1.27 $\mu$m \ce{O2} $^{1}\Delta_{g}$ band---which is suppressed by approximately a factor of 200 for the 10-bar \ce{O2} atmospheres when compared to the 1 bar Earth-like atmosphere---and least pronounced in the 0.69 $\mu$m \ce{O2} B-band, which is suppressed by a factor of \~3.    

The different impact of the O$_2$-O$_2$ CIA on the \ce{O2} bands affords a potential, and critical test for the biogenicity of the observed \ce{O2}.  Because of the high suppression factor in the presence of very high levels of \ce{O2}, the 1.27 $\mu$m \ce{O2} $^{1}\Delta_{g}$ band is the most sensitive spectral indicator for the relatively small amounts of \ce{O2} expected from a biosphere. If the 1.27 $\mu$m band is detected, the atmospheric \ce{O2} is more likely to be due to life, and it is unlikely that a massive \ce{O2} atmosphere is present. In comparison, if the 0.69 $\mu$m \ce{O2} B-band (or 0.76 $\mu$m A-band) is detected, but the (normally much stronger) 1.27 $\mu$m band is not, this would point towards a more massive atmosphere, with ocean loss as the more likely source of any \ce{O2} detected.  

Observing multiple \ce{O2} bands for a given target will be challenging, but likely feasible with upcoming high-resolution spectrometers planned for ground-based ELTs, including the potential detection of the 1.27 $\mu$m band \citep{Kawahara2012}. Our results confirm previous work \citep{Snellen2013} on the relatively enhanced detectability of the 1.27 $\mu$m with respect to the 0.76 $\mu$m band for this M5.5V host star, although we did not consider the effect of correlated ``red'' noise, which may suppress the detectability of the 1.27 $\mu$m band for transmission spectra  \citep{Lopez-MoralesMercedes2019DEBo}. Additionally, recent work by \citet{Serindag2019} using injection of a simulated Earth-like transmission spectrum into high-resolution observations of Proxima Centauri suggests that neglecting correlated noise may not significantly alter the detectability of the 0.76 $\mu$m band when real stellar data are used.  Calculations of the detectability of the 0.76 $\mu$m \ce{O2} band in reflected light observations for spectrometers with spectral resolution up to $5 \times 10 ^ 5$ suggest that these observations are likely feasible in 10s of hours of observing time    \citep{Lovis2017,HawkerPerry2019}.

In the near-term, high-resolution ground-based observations are therefore likely the best, if not the only, way to detect biogenic levels of \ce{O2}.  These studies will complement near-term space-based low-resolution observations of a handful of transiting HZ targets, which will be more sensitive to signatures of ocean-loss, rather than photosynthesis. In particular, JWST may be able to use transmission observations to detect strong absorption from \ce{O2-O2} CIA in as few as 4-20 transits of the planets TRAPPIST-1 b-e \citep{Lincowski2018,Lustig-Yaeger2019}, if these exoplanets have \ce{O2}-dominated post-ocean-loss atmospheres.  However, detection of biogenic levels of \ce{O2} on TRAPPIST-1 e would require well over 100 transits, and is likely infeasible with JWST \citep{Lustig-Yaeger2019,Wunderlich2019detectability}. While our simulations show that the \ce{O2}-\ce{O2} CIA suppresses the signal for ground-based cross-correlation techniques, there may be future opportunities to explore a positive detection of the \ce{O2}-\ce{O2} feature in ground-based spectra using reflected stellar lines to measure wavelength-dependent albedo differences and potentially trace this broad absorption. 

\section{Conclusion}
\label{sec:conclusion}

In the next decade, ground-based extremely large telescopes will use high-resolution spectrographs to search for \ce{O2} in the atmospheres of terrestrial exoplanets. Discriminating ocean loss planets is a critical step in assessing planetary habitability and determining whether or not any O$_2$ detected in a planet's atmosphere is due to life. We have used self-consistent coupled climate-photochemical models of an Earth-like planet orbiting Proxima Centauri and compared this with the spectrum of planet that has suffered ocean loss and retained a 10-bar O$_2$ atmosphere. Suppression of the continuum near oxygen bands as a result of O$_2$-O$_2$ CIA is a key feature of high-O$_2$ atmospheres. Our results show that any detection of the 1.27$\mu$m band in reflected light suggests that the planet has the lower Earth-like \ce{O2} levels that are more likely to be indicative of photosynthesis.  Conversely, because the CIA suppression does not have a strong effect at the 0.69 $\mu$m oxygen B band, detection of the 0.69 $\mu$m oxygen B band, or the 0.76 $\mu$m A-band, and a non-detection or strong suppression of the 1.27 $\mu$m band would suggest ocean loss as the likely source of the observed O$_2$. 

\acknowledgements
We thank an anonymous reviewer for an extremely helpful and generous review and comments that greatly improved the quality of this manuscript, including the idea of detecting CIA using reflected stellar light.
This work was supported by an award from the National Institute of Natural Sciences, Japan, administered via the Astrobiology Center at the University of Tokyo.  
We also acknowledge support from NASA's NExSS Virtual Planetary Laboratory, funded under NASA Astrobiology Institute Cooperative Agreement Number NNA13AA93A, and the NASA Astrobiology Program grant 80NSSC18K0829.  
This work benefited from participation in the NASA Nexus for Exoplanet Systems Science research coordination network. This work made use of the advanced computational, storage, and networking infrastructure provided by the Hyak supercomputer system at the University of Washington.

\software{SMART \citep{Meadows1996}, Matplotlib \citep{Hunter2007}, NumPy \citep{Walt2011}, SciPy \citep{Virtanen2020}}

\bibliographystyle{aasjournal}
\bibliography{references}

\begin{thebibliography}{}
\expandafter\ifx\csname natexlab\endcsname\relax\def\natexlab#1{#1}\fi
\providecommand{\url}[1]{\href{#1}{#1}}
\providecommand{\dodoi}[1]{doi:~\href{http://doi.org/#1}{\nolinkurl{#1}}}
\providecommand{\doeprint}[1]{\href{http://ascl.net/#1}{\nolinkurl{http://ascl.net/#1}}}
\providecommand{\doarXiv}[1]{\href{https://arxiv.org/abs/#1}{\nolinkurl{https://arxiv.org/abs/#1}}}

\bibitem[{Anglada-Escud{\'e} {et~al.}(2016)Anglada-Escud{\'e}, Amado, Barnes,
  Berdi{\~n}as, Butler, Coleman, de~La~Cueva, Dreizler, Endl, Giesers,
  {et~al.}}]{Anglada2016}
Anglada-Escud{\'e}, G., Amado, P.~J., Barnes, J., {et~al.} 2016, Nature, 536,
  437

\bibitem[{Arney {et~al.}(2014)Arney, Meadows, Crisp, Schmidt, Bailey, \&
  Robinson}]{Arney2014}
Arney, G., Meadows, V., Crisp, D., {et~al.} 2014, Journal of Geophysical
  Research: Planets, 119, 1860, \dodoi{10.1002/2014JE004662}

\bibitem[{Baraffe {et~al.}(2015)Baraffe, Homeier, Allard, \&
  Chabrier}]{Baraffe2015}
Baraffe, I., Homeier, D., Allard, F., \& Chabrier, G. 2015, Astronomy \&
  Astrophysics, 577, A42

\bibitem[{Barnes {et~al.}(2013)Barnes, Mullins, Goldblatt, Meadows, Kasting, \&
  Heller}]{Barnes2013a}
Barnes, R., Mullins, K., Goldblatt, C., {et~al.} 2013, Astrobiology, 13, 225,
  \dodoi{10.1089/ast.2012.0851}

\bibitem[{Berta-Thompson {et~al.}(2015)Berta-Thompson, Irwin, Charbonneau,
  Newton, Dittmann, Astudillo-Defru, Bonfils, Gillon, Jehin, Stark,
  {et~al.}}]{Berta2015}
Berta-Thompson, Z.~K., Irwin, J., Charbonneau, D., {et~al.} 2015, Nature, 527,
  204

\bibitem[{{Bonfils, X.} {et~al.}(2018){Bonfils, X.}, {Astudillo-Defru, N.},
  {D\'{\i}az, R.}, {Almenara, J.-M.}, {Forveille, T.}, {Bouchy, F.}, {Delfosse,
  X.}, {Lovis, C.}, {Mayor, M.}, {Murgas, F.}, {Pepe, F.}, {Santos, N. C.},
  {S\'egransan, D.}, {Udry, S.}, \& {W\"unsche, A.}}]{Bonfils2018}
{Bonfils, X.}, {Astudillo-Defru, N.}, {D\'{\i}az, R.}, {et~al.} 2018, A\&A,
  613, A25, \dodoi{10.1051/0004-6361/201731973}

\bibitem[{Chassefi{\`e}re(1997)}]{Chassefiere1997}
Chassefi{\`e}re, E. 1997, Icarus, 126, 229

\bibitem[{Del~Genio {et~al.}(2019)Del~Genio, Way, Amundsen, Aleinov, Kelley,
  Kiang, \& Clune}]{DelGenio2019Proxima}
Del~Genio, A.~D., Way, M.~J., Amundsen, D.~S., {et~al.} 2019, Astrobiology, 19,
  99

\bibitem[{Dittmann {et~al.}(2017)Dittmann, Irwin, Charbonneau, Bonfils,
  Astudillo-Defru, Haywood, Berta-Thompson, Newton, Rodriguez, Winters,
  {et~al.}}]{Dittmann2017}
Dittmann, J.~A., Irwin, J.~M., Charbonneau, D., {et~al.} 2017, Nature, 544, 333

\bibitem[{Domagal-Goldman {et~al.}(2014)Domagal-Goldman, Segura, Claire,
  Robinson, \& Meadows}]{Domagal-Goldman2014}
Domagal-Goldman, S.~D., Segura, A., Claire, M.~W., Robinson, T.~D., \& Meadows,
  V.~S. 2014, The Astrophysical Journal, 792, 90

\bibitem[{Dotter {et~al.}(2008)Dotter, Chaboyer, Jevremovi, Kostov, Baron, \&
  Ferguson}]{Dotter2008}
Dotter, A., Chaboyer, B., Jevremovi, D., {et~al.} 2008, The Astrophysical
  Journal Supplement Series, 178, 89

\bibitem[{Dressing \& Charbonneau(2015)}]{Dressing2015}
Dressing, C.~D., \& Charbonneau, D. 2015, The Astrophysical Journal, 807, 45

\bibitem[{{Erkaev} {et~al.}(2007){Erkaev}, {Kulikov}, {Lammer}, {Selsis},
  {Langmayr}, {Jaritz}, \& {Biernat}}]{Erkaev2007}
{Erkaev}, N.~V., {Kulikov}, Y.~N., {Lammer}, H., {et~al.} 2007, A\&A, 472, 329,
  \dodoi{10.1051/0004-6361:20066929}

\bibitem[{Froning {et~al.}(2006)Froning, Osterman, Beasley, Green, \&
  Beland}]{Froning2006}
Froning, C., Osterman, S., Beasley, M., Green, J., \& Beland, S. 2006, in
  Ground-based and Airborne Instrumentation for Astronomy, ed. I.~S. McLean \&
  M.~Iye, Vol. 6269, International Society for Optics and Photonics (SPIE), 626
  -- 634.
\newblock \url{https://doi.org/10.1117/12.669358}

\bibitem[{Gebauer {et~al.}(2018)Gebauer, Grenfell, Lehmann, \&
  Rauer}]{Gebauer2018}
Gebauer, S., Grenfell, J.~L., Lehmann, R., \& Rauer, H. 2018, Astrobiology, 18

\bibitem[{Gillon {et~al.}(2017)Gillon, Triaud, Demory, Jehin, Agol, Deck,
  Lederer, De~Wit, Burdanov, Ingalls, {et~al.}}]{Gillon2017}
Gillon, M., Triaud, A.~H., Demory, B.-O., {et~al.} 2017, Nature, 542, 456

\bibitem[{{Gordon} {et~al.}(2017){Gordon}, {Rothman}, {Hill}, {Kochanov},
  {Tan}, {Bernath}, {Birk}, {Boudon}, {Campargue}, {Chance}, {Drouin}, {Flaud},
  {Gamache}, {Hodges}, {Jacquemart}, {Perevalov}, {Perrin}, {Shine}, {Smith},
  {Tennyson}, {Toon}, {Tran}, {Tyuterev}, {Barbe}, {Cs{\'a}sz{\'a}r}, {Devi},
  {Furtenbacher}, {Harrison}, {Hartmann}, {Jolly}, {Johnson}, {Karman},
  {Kleiner}, {Kyuberis}, {Loos}, {Lyulin}, {Massie}, {Mikhailenko},
  {Moazzen-Ahmadi}, {M{\"u}ller}, {Naumenko}, {Nikitin}, {Polyansky}, {Rey},
  {Rotger}, {Sharpe}, {Sung}, {Starikova}, {Tashkun}, {Auwera}, {Wagner},
  {Wilzewski}, {Wcis{\l}o}, {Yu}, \& {Zak}}]{Gordon2017}
{Gordon}, I.~E., {Rothman}, L.~S., {Hill}, C., {et~al.} 2017, \jqsrt, 203, 3,
  \dodoi{10.1016/j.jqsrt.2017.06.038}

\bibitem[{Hawker \& Parry(2019)}]{HawkerPerry2019}
Hawker, G.~A., \& Parry, I.~R. 2019, MNRAS

\bibitem[{{Hunter}(2007)}]{Hunter2007}
{Hunter}, J.~D. 2007, Computing in Science Engineering, 9, 90

\bibitem[{Karman {et~al.}(2019)Karman, Gordon, van Der~Avoird, Baranov, Boulet,
  Drouin, Groenenboom, Gustafsson, Hartmann, Kurucz, Rothman, Sun, Sung,
  Thalman, Tran, Wishnow, Wordsworth, Vigasin, Volkamer, \& van
  Der~Zande}]{Karman2019}
Karman, T., Gordon, I.~E., van Der~Avoird, A., {et~al.} 2019, Icarus, 328, 160

\bibitem[{Kawahara {et~al.}(2012)Kawahara, Matsuo, Takami, Fujii, Kotani,
  Murakami, Tamura, \& Guyon}]{Kawahara2012}
Kawahara, H., Matsuo, T., Takami, M., {et~al.} 2012, The Astrophysical Journal,
  758, 13, \dodoi{10.1088/0004-637x/758/1/13}

\bibitem[{Kleine {et~al.}(2009)Kleine, Touboul, Bourdon, Nimmo, Mezger, Palme,
  Jacobsen, Yin, \& Halliday}]{Kleine2009}
Kleine, T., Touboul, M., Bourdon, B., {et~al.} 2009, Geochimica et Cosmochimica
  Acta, 73, 5150

\bibitem[{Kopparapu {et~al.}(2013)Kopparapu, Ramirez, Kasting, Eymet, Robinson,
  Mahadevan, Terrien, Domagal-Goldman, Meadows, \& Deshpande}]{Kopparapu2013a}
Kopparapu, R.~K., Ramirez, R., Kasting, J.~F., {et~al.} 2013, The Astrophysical
  Journal, 765, 16, \dodoi{10.1088/0004-637X/765/2/131}

\bibitem[{{Lee} {et~al.}(2010){Lee}, {Yuk}, {Lee}, {Wang}, {Park}, {Park},
  {Chun}, {Pak}, {Strubhar}, {Deen}, {Gully-Santiago}, {Rand}, {Seo}, {Kwon},
  {Oh}, {Barnes}, {Lacy}, {Goertz}, {Park}, {Pyo}, \& {Jaffe}}]{Lee2010}
{Lee}, S., {Yuk}, I.-S., {Lee}, H., {et~al.} 2010, Society of Photo-Optical
  Instrumentation Engineers (SPIE) Conference Series, Vol. 7735, {GMTNIRS
  (Giant Magellan Telescope near-infrared spectrograph): design concept},
  77352K

\bibitem[{Lincowski {et~al.}(2018)Lincowski, Meadows, Crisp, Robinson, Luger,
  Lustig-Yaeger, \& Arney}]{Lincowski2018}
Lincowski, A.~P., Meadows, V.~S., Crisp, D., {et~al.} 2018, The Astrophysical
  Journal, 867, 76

\bibitem[{{L{\'{o}}pez-Morales} {et~al.}(2019){L{\'{o}}pez-Morales}, {Currie},
  {Teske}, {Gaidos}, {Kempton}, {Males}, {Lewis}, {Rackham}, {Ben-Ami},
  {Birkby}, {Charbonneau}, {Close}, {Crane}, {Dressing}, {Froning}, {Hasegawa},
  {Konopacky}, {Kopparapu}, {Mawet}, {Mennesson}, {Ramirez}, {Stelter},
  {Szentgyorgyi}, {Wang}, {Alam}, {Collins}, {Dupree}, {Karovska}, {Kirk},
  {Levi}, {McGruder}, {Packman}, {Rugheimer}, \&
  {Rukdee}}]{Lopez-MoralesMercedes2019DEBo}
{L{\'{o}}pez-Morales}, M., {Currie}, T., {Teske}, J., {et~al.} 2019, \baas, 51,
  162.
\newblock \doarXiv{1903.09523}

\bibitem[{{Lovis} {et~al.}(2017){Lovis}, {Snellen}, {Mouillet}, {Pepe},
  {Wildi}, {Astudillo-Defru}, {Beuzit}, {Bonfils}, {Cheetham}, {Conod},
  {Delfosse}, {Ehrenreich}, {Figueira}, {Forveille}, {Martins}, {Quanz},
  {Santos}, {Schmid}, {S{\'e}gransan}, \& {Udry}}]{Lovis2017}
{Lovis}, C., {Snellen}, I., {Mouillet}, D., {et~al.} 2017, \aap, 599, A16,
  \dodoi{10.1051/0004-6361/201629682}

\bibitem[{Luger {et~al.}(2015)Luger, Barnes, Lopez, Fortney, Jackson, \&
  Meadows}]{Luger2015a}
Luger, R., Barnes, R., Lopez, E., {et~al.} 2015, Astrobiology, 15, 57,
  \dodoi{10.1089/ast.2014.1215}

\bibitem[{{Lustig-Yaeger} {et~al.}(2019){Lustig-Yaeger}, {Meadows}, \&
  {Lincowski}}]{Lustig-Yaeger2019}
{Lustig-Yaeger}, J., {Meadows}, V.~S., \& {Lincowski}, A.~P. 2019, \aj, 158,
  27, \dodoi{10.3847/1538-3881/ab21e0}

\bibitem[{Marconi {et~al.}(2016)Marconi, Marcantonio, D'Odorico, Cristiani,
  Maiolino, Oliva, Origlia, Riva, Valenziano, Zerbi, Abreu, Adibekyan, Prieto,
  Amado, Benz, Boisse, Bonfils, Bouchy, Buchhave, Buscher, Cabral, Martins,
  Chiavassa, Coelho, Christensen, Delgado-Mena, de~Medeiros, Varano, Figueira,
  Fisher, Fynbo, Glasse, Haehnelt, Haniff, Hansen, Hatzes, Huke, Korn, Leão,
  Liske, Lovis, Maslowski, Matute, McCracken, Martins, Monteiro, Morris,
  Morris, Nicklas, Niedzielski, Nunes, Palle, Parr-Burman, Parro, Parry, Pepe,
  Piskunov, Queloz, Quirrenbach, Lopez, Reiners, Reid, Santos, Seifert, Sousa,
  Stempels, Strassmeier, Sun, Udry, Vanzi, Vestergaard, Weber, \&
  Zackrisson}]{Marconi2016}
Marconi, A., Marcantonio, P.~D., D'Odorico, V., {et~al.} 2016, in Ground-based
  and Airborne Instrumentation for Astronomy VI, ed. C.~J. Evans, L.~Simard, \&
  H.~Takami, Vol. 9908, International Society for Optics and Photonics (SPIE),
  676 -- 687.
\newblock \url{https://doi.org/10.1117/12.2231653}

\bibitem[{Meadows(2017)}]{Meadows2017a}
Meadows, V.~S. 2017, Astrobiology

\bibitem[{{Meadows} \& {Crisp}(1996)}]{Meadows1996}
{Meadows}, V.~S., \& {Crisp}, D. 1996, \jgr, 101, 4595,
  \dodoi{10.1029/95JE03567}

\bibitem[{Meadows {et~al.}(2018{\natexlab{a}})Meadows, Reinhard, Arney,
  Parenteau, Schwieterman, Domagal-Goldman, Lincowski, Stapelfeldt, Rauer,
  DasSarma, Hegde, Narita, Deitrick, Lustig-Yaeger, Lyons, Siegler, \&
  Grenfell}]{Meadows2018b}
Meadows, V.~S., Reinhard, C.~T., Arney, G.~N., {et~al.} 2018{\natexlab{a}},
  Astrobiology, 18, 630, \dodoi{10.1089/ast.2017.1727}

\bibitem[{Meadows {et~al.}(2018{\natexlab{b}})Meadows, Arney, Schwieterman,
  Lustig-Yaeger, Lincowski, Robinson, Domagal-Goldman, Deitrick, Barnes,
  Fleming, Luger, Driscoll, Quinn, \& Crisp}]{Meadows2018a}
Meadows, V.~S., Arney, G.~N., Schwieterman, E.~W., {et~al.} 2018{\natexlab{b}},
  Astrobiology, 18, 133, \dodoi{10.1089/ast.2016.1589}

\bibitem[{{Ribas} {et~al.}(2016){Ribas}, {Bolmont}, {Selsis}, {Reiners},
  {Leconte}, {Raymond}, {Engle}, {Guinan}, {Morin}, {Turbet}, {Forget}, \&
  {Anglada-Escud{\'e}}}]{Ribas2016}
{Ribas}, I., {Bolmont}, E., {Selsis}, F., {et~al.} 2016, \aap, 596, A111,
  \dodoi{10.1051/0004-6361/201629576}

\bibitem[{{Robinson} {et~al.}(2016){Robinson}, {Stapelfeldt}, \&
  {Marley}}]{Robinson2016}
{Robinson}, T.~D., {Stapelfeldt}, K.~R., \& {Marley}, M.~S. 2016, \pasp, 128,
  025003, \dodoi{10.1088/1538-3873/128/960/025003}

\bibitem[{Robinson {et~al.}(2011)Robinson, Meadows, Crisp, Deming, A'Hearn,
  Charbonneau, Livengood, Seager, Barry, Hearty, {et~al.}}]{Robinson2011}
Robinson, T.~D., Meadows, V.~S., Crisp, D., {et~al.} 2011, Astrobiology, 11,
  393

\bibitem[{{Rodler} \& {L{\'o}pez-Morales}(2014)}]{Rodler2014}
{Rodler}, F., \& {L{\'o}pez-Morales}, M. 2014, \apj, 781, 54,
  \dodoi{10.1088/0004-637X/781/1/54}

\bibitem[{Schaefer {et~al.}(2016)Schaefer, Wordsworth, Berta-Thompson, \&
  Sasselov}]{Schaefer2016}
Schaefer, L., Wordsworth, R.~D., Berta-Thompson, Z., \& Sasselov, D. 2016, The
  Astrophysical Journal, 829, 63

\bibitem[{{Schwieterman} {et~al.}(2016){Schwieterman}, {Meadows},
  {Domagal-Goldman}, {Deming}, {Arney}, {Luger}, {Harman}, {Misra}, \&
  {Barnes}}]{Schwieterman2016}
{Schwieterman}, E.~W., {Meadows}, V.~S., {Domagal-Goldman}, S.~D., {et~al.}
  2016, The Astrophysical Journal Letters, 819, L13,
  \dodoi{10.3847/2041-8205/819/1/L13}

\bibitem[{{Segura} {et~al.}(2010){Segura}, {Walkowicz}, {Meadows}, {Kasting},
  \& {Hawley}}]{Segura2010}
{Segura}, A., {Walkowicz}, L.~M., {Meadows}, V., {Kasting}, J., \& {Hawley}, S.
  2010, Astrobiology, 10, 751, \dodoi{10.1089/ast.2009.0376}

\bibitem[{Serindag \& Snellen(2019)}]{Serindag2019}
Serindag, D.~B., \& Snellen, I. A.~G. 2019, The Astrophysical Journal, 871, L7,
  \dodoi{10.3847/2041-8213/aafa1f}

\bibitem[{Shields {et~al.}(2016)Shields, Barnes, Agol, Charnay, Bitz, \&
  Meadows}]{Shields2016b}
Shields, A.~L., Barnes, R., Agol, E., {et~al.} 2016, Astrobiology, 16

\bibitem[{{Snellen} {et~al.}(2015){Snellen}, {de Kok}, {Birkby}, {Brandl},
  {Brogi}, {Keller}, {Kenworthy}, {Schwarz}, \& {Stuik}}]{Snellen2015}
{Snellen}, I., {de Kok}, R., {Birkby}, J.~L., {et~al.} 2015, \aap, 576, A59,
  \dodoi{10.1051/0004-6361/201425018}

\bibitem[{{Snellen} {et~al.}(2013){Snellen}, {de Kok}, {le Poole}, {Brogi}, \&
  {Birkby}}]{Snellen2013}
{Snellen}, I.~A.~G., {de Kok}, R.~J., {le Poole}, R., {Brogi}, M., \& {Birkby},
  J. 2013, \apj, 764, 182, \dodoi{10.1088/0004-637X/764/2/182}

\bibitem[{{Snellen} {et~al.}(2017){Snellen}, {D{\'e}sert}, {Waters},
  {Robinson}, {Meadows}, {van Dishoeck}, {Brandl}, {Henning}, {Bouwman},
  {Lahuis}, {Min}, {Lovis}, {Dominik}, {Van Eylen}, {Sing},
  {Anglada-Escud{\'e}}, {Birkby}, \& {Brogi}}]{Snellen2017}
{Snellen}, I.~A.~G., {D{\'e}sert}, J.-M., {Waters}, L.~B.~F.~M., {et~al.} 2017,
  \aj, 154, 77, \dodoi{10.3847/1538-3881/aa7fbc}

\bibitem[{{Szentgyorgyi} {et~al.}(2016){Szentgyorgyi}, {Baldwin}, {Barnes},
  {Bean}, {Ben-Ami}, {Brennan}, {Budynkiewicz}, {Chun}, {Conroy}, {Crane},
  {Epps}, {Evans}, {Evans}, {Foster}, {Frebel}, {Gauron}, {Guzm{\'a}n}, {Hare},
  {Jang}, {Jang}, {Jordan}, {Kim}, {Kim}, {Mendes de Oliveira},
  {Lopez-Morales}, {McCracken}, {McMuldroch}, {Miller}, {Mueller}, {Oh},
  {Onyuksel}, {Ordway}, {Park}, {Park}, {Park}, {Paxson}, {Phillips},
  {Plummer}, {Podgorski}, {Seifahrt}, {Stark}, {Steiner}, {Uomoto},
  {Walsworth}, \& {Yu}}]{Szentgyorgyi2016}
{Szentgyorgyi}, A., {Baldwin}, D., {Barnes}, S., {et~al.} 2016, in \procspie,
  Vol. 9908, Ground-based and Airborne Instrumentation for Astronomy VI, 990822

\bibitem[{Tinetti {et~al.}(2005)Tinetti, Meadows, Crisp, Fong, Velusamy, \&
  Snively}]{Tinetti2005}
Tinetti, G., Meadows, V.~S., Crisp, D., {et~al.} 2005, Astrobiology, 5, 461,
  \dodoi{10.1089/ast.2005.5.461}

\bibitem[{{van der Walt} {et~al.}(2011){van der Walt}, {Colbert}, \&
  {Varoquaux}}]{Walt2011}
{van der Walt}, S., {Colbert}, S.~C., \& {Varoquaux}, G. 2011, Computing in
  Science Engineering, 13, 22

\bibitem[{Virtanen {et~al.}(2020)Virtanen, Gommers, Oliphant, Haberland, Reddy,
  Cournapeau, Burovski, Peterson, Weckesser, Bright, van~der Walt, Brett,
  Wilson, Millman, Mayorov, Nelson, Jones, Kern, Larson, \&
  Carey}]{Virtanen2020}
Virtanen, P., Gommers, R., Oliphant, T., {et~al.} 2020, Nature Methods, 17

\bibitem[{Wunderlich {et~al.}(2019)Wunderlich, Godolt, Grenfell, St{\"a}dt,
  Smith, Gebauer, Schreier, Hedelt, \& Rauer}]{Wunderlich2019detectability}
Wunderlich, F., Godolt, M., Grenfell, J.~L., {et~al.} 2019, Astronomy \&
  Astrophysics, 624, A49

\bibitem[{Yang {et~al.}(2017)Yang, Wang, Baade, Brown, Cracraft, Hoflich,
  Maund, Patat, Sparks, Spyromilio, {et~al.}}]{Yang2017}
Yang, Y., Wang, L., Baade, D., {et~al.} 2017, arXiv preprint:1704.01431

\bibitem[{Young(1980)}]{Young1980}
Young, A.~T. 1980, Appl. Opt., 19, 3427, \dodoi{10.1364/AO.19.003427}

\end{thebibliography}

\end{document}